\documentclass{article}

\usepackage{arxiv}

\usepackage[utf8]{inputenc} 
\usepackage[T1]{fontenc}    
\usepackage{hyperref}       
\usepackage{url}            
\usepackage{booktabs}       
\usepackage{amsfonts}       
\usepackage{nicefrac}       
\usepackage{graphicx}
\usepackage[numbers,sort&compress]{natbib}
\usepackage{doi}
\usepackage{amsmath}        
\usepackage{cleveref}       
\usepackage{microtype}      
\usepackage{multirow}
\usepackage{tabularx}
\usepackage{chngcntr}
\usepackage{changes}

\title{Retrospective macromolecular pool fraction mapping from conventional clinical MRI: a physics-guided self-supervised deep learning framework}

\date{}

\newif\ifuniqueAffiliation

\ifuniqueAffiliation 
\else
\usepackage{authblk}

\author[1]{Jelmer van Lune}
\author[1]{Miha Fuderer}
\author[1]{Oscar van der Heide}
\author[1]{Matteo Maspero}
\author[1]{Martin B.~Schilder}
\author[2]{Jan Willem Dankbaar}
\author[1]{Stefano Mandija}
\author[1]{Alessandro Sbrizzi}

\affil[1]{Computational Imaging Group for MRI Therapy \& Diagnostics, University Medical Center Utrecht, Utrecht, The Netherlands}
\affil[2]{Department of Radiology, University Medical Center Utrecht, Utrecht, The Netherlands}
\fi

\renewcommand{\shorttitle}{}

\hypersetup{
pdftitle={},
pdfauthor={Jelmer van Lune},
}

\begin{document}
\maketitle

\begin{abstract}
Macromolecular pool fraction (MPF) is a quantitative imaging biomarker sensitive to myelin integrity and other changes in tissue macromolecular content, but its dependence on specialized acquisitions and reconstruction methods limits its availability for large-scale studies. We present a physics-guided, self-supervised deep learning framework for retrospectively estimating MPF directly from conventional clinical brain MRI contrasts ($T_1$-weighted, $T_2$-weighted, and fluid-attenuated inversion recovery [FLAIR]), without paired quantitative reference scans. The framework embeds an analytical two-pool magnetization transfer signal model that relates sequence-averaged radiofrequency transmit power to apparent longitudinal relaxation. To accommodate heterogeneous clinical protocols, the network conditions intermediate feature representations on acquisition parameters via adaptive instance normalization. The framework was trained and evaluated on a multisite, multivendor, multiprotocol dataset comprising 7,705 MRI sessions (3~T) from 4,489 subjects across Philips and Siemens systems. Across 1,237 test sessions, mean MPF was $0.182 \pm 0.007$ in white matter and $0.144 \pm 0.009$ in gray matter, with limited variation across vendor-by-protocol groups (between-group coefficient of variation: 2.65\% in white matter and 5.63\% in gray matter). In 70 sessions with demyelinating-disease labels, MPF was 35.6\% lower in segmented lesions than in normal-appearing white matter. As a technical demonstration, MPF maps from two different clinical protocols acquired in a healthy volunteer showed a mean voxel-wise absolute percentage difference of 7.7\% and a concordance correlation coefficient of $0.888$, indicating good cross-protocol consistency. These results suggest that magnetization transfer-informed self-supervised learning enables stable estimation of MPF maps from conventional clinical MRI, providing a scalable approach to investigate demyelinating pathology in both existing imaging archives and prospectively acquired conventional clinical scans without dedicated qMT acquisitions.
\end{abstract}



\section{Introduction}
Magnetic resonance imaging (MRI) provides excellent soft-tissue contrast and is routinely used for clinical diagnosis. Signal intensity in conventional MR images depends on tissue properties such as longitudinal and transverse relaxation times ($T_1$, $T_2$, $T_2^*$) and proton-density (PD). In addition, magnetization transfer (MT) is a fundamental tissue- and sequence-dependent effect that influences the MR signal~\cite{Wolff1989}. MT is commonly described using a two-pool model comprising a free-water pool and a semisolid pool~\cite{Henkelman1993}. The semisolid pool contains protons bound to macromolecules, including proteins and lipids, and has a transverse relaxation time on the order of microseconds. It is therefore not directly observable with conventional MRI. However, exchange of longitudinal magnetization between the free and semisolid pools alters the longitudinal relaxation of the observable free-water signal and produces biexponential relaxation. MT is consequently a major contributor to longitudinal relaxation in biological tissue~\cite{Asslnder2024a}. The relative size of the semisolid pool is quantified by the macromolecular pool fraction (MPF, denoted $m_0^\mathrm{s}$), which quantifies the relative amount of protons bound to biological macromolecules that exchange magnetization with water protons~\cite{Kisel2022}.


Among quantitative MT tissue parameters, MPF has attracted particular interest as a marker of myelin integrity in multiple sclerosis (MS), the most common demyelinating disease of the central nervous system~\cite{Wolff1994, Henkelman2001, Sled2001}. MPF is sensitive to myelin content and is markedly reduced in MS lesions~\cite{Henkelman2001, Asslnder2024a, Kisel2022, Sled2018, Levesque2010, Yarnykh2002}. This sensitivity has enabled investigations of spatial heterogeneity in demyelination and of lesion evolution over time~\cite{Kisel2022, Levesque2005, Levesque2010, Clarke2021, Giacomini2009}. Moreover, post-mortem studies associated reduced MPF with histopathological evidence of demyelination~\cite{Kisel2022, Schmierer2007, Bagnato2018}. MPF mapping could therefore provide a quantitative imaging readout for studying demyelinating pathology, monitoring disease progression, and assessing treatment response.

Despite this clinical potential, MPF mapping requires dedicated prospective quantitative MT (qMT) acquisitions with specialized pulse sequences and reconstruction pipelines that are generally not included in standard neuroimaging protocols. Several qMT methods have been developed over the past three decades, including multi-point off-resonance Z-spectroscopy~\cite{Henkelman1993, Sled2001, Yarnykh2002}, rapid single-point mapping~\cite{Yarnykh2012, Yarnykh2016}, on-resonance selective inversion recovery~\cite{Gochberg2003, Dortch2011}, steady-state saturation~\cite{Gloor2008}, and transient hybrid-state methods~\cite{Asslnder2019, Asslnder2024a, Asslnder2024b}. These approaches require additional acquisition time and can only be applied prospectively. In addition, they cannot recover MPF from the large collections of conventional scans that already exist in clinical archives. These limitations restrict access to MPF maps and impede both large-scale biomarker studies and subsequent clinical translation.


The widespread availability of routine clinical MRI provides an opportunity to derive quantitative tissue parameters without dedicated prospective acquisitions. Initial efforts to exploit these routine scans have relied on mathematical combinations of conventional contrasts—such as $T_1\mathrm{w}/T_2\mathrm{w}$ and $T_1\mathrm{w}/\mathrm{FLAIR}$ ratios or empirical polynomial fits—as accessible proxies for myelin integrity and myelin water fraction (MWF)~\cite{Ganzetti2014, Cappelle2022, Drenthen2026}. However, these approaches provide relative, intensity-based indices rather than quantitative physical parameters. 

More recently, self-supervised, physics-guided deep learning frameworks have begun to address these limitations by retrospectively estimating quantitative maps from routine multi-contrast acquisitions ($T_1\mathrm{w}$, $T_2\mathrm{w}$, and FLAIR) without requiring paired reference scans~\cite{vanLune2026, Qiu2024, Moya-Sez2021ISMRM}. These frameworks embed analytical Bloch equations-based signal models directly into the training objective, enabling scalable deployment across clinical archives. Existing frameworks, however, assume single-pool relaxation and remain restricted to relaxometry parameters ($T_1$ and $T_2$) alongside PD. Consequently, the retrospective estimation of quantitative MT parameters, including the MPF, from conventional clinical MRI remains unexplored.

In this work, we introduce a framework for estimating an MT-model-based MPF directly from routine $T_1\mathrm{w}$, $T_2\mathrm{w}$, and FLAIR images without paired qMT reference data. We extended a physics-guided, self-supervised deep learning framework by incorporating a simplified closed-form description of MT into its forward signal model. Following Teixeira et al.~\cite{Rui2019}, the model relates the MT-induced apparent longitudinal relaxation time to the tissue-dependent MPF and the average RF transmit power, $B_{1,\mathrm{rms}}$, the latter being determined by the acquisition protocol. To reduce protocol-dependent variation in these estimated maps, we conditioned the network on the input scans' acquisition parameters. We trained and evaluated the framework using a clinically heterogeneous, multicenter, multivendor dataset comprising more than 7,700 MRI sessions (23,000+ image volumes), including sessions from patients with demyelinating lesions. As a technical demonstration, we assessed cross-protocol consistency using repeated acquisitions in a healthy volunteer.

\section{Theory}
\label{sec:theory}

\subsection{Two-pool magnetization transfer model} 

MT in biological tissue is generally modeled using a two-pool framework comprising a liquid free-water pool (superscript $\mathrm{f}$) and a semisolid macromolecular pool (superscript $\mathrm{s}$)~\cite{Henkelman1993}. Each pool is characterized by its own longitudinal and transverse relaxation rates. However, the transverse relaxation time of the semisolid pool, $T_2^\mathrm{s}$, is on the order of microseconds, preventing its direct observation with conventional MRI. Longitudinal magnetization is exchanged between the two pools with first-order rate constants $k^\mathrm{f}$ (free to semisolid) and $k^\mathrm{s}$ (semisolid to free)~\cite{Rui2019}.

In this work, we define the macromolecular pool fraction (MPF) as the fractional equilibrium magnetization of the semisolid pool relative to the total equilibrium magnetization. Here, $M_0^\mathrm{f}$ and $M_0^\mathrm{s}$ denote the equilibrium longitudinal magnetizations of the free and semisolid pools, respectively. The MPF is therefore defined as $m_0^\mathrm{s} = M_0^\mathrm{s}/(M_0^\mathrm{f} + M_0^\mathrm{s})$, such that $m_0^\mathrm{f} + m_0^\mathrm{s} = 1$. Note that this definition follows recent quantitative MT literature~\cite{Yarnykh2012, Asslnder2024a, Asslnder2024b, Asslnder2025var}. In contrast, earlier qMT studies~\cite{Henkelman1993, Gochberg2003, Sled2001} frequently report the pool size ratio relative to the free pool, $F = M_0^\mathrm{s}/M_0^\mathrm{f}$. The two definitions are related by 

\begin{equation} m_0^\mathrm{s} = \frac{F}{1 + F} \quad \Longleftrightarrow \quad F = \frac{m_0^\mathrm{s}}{1 - m_0^\mathrm{s}}.
\label{eq:mpf_conversion}
\end{equation}

Following~\citet{Asslnder2024a}, we express both rate constants through a single exchange rate $R_x$ as $k^\mathrm{f} = R_x m_0^\mathrm{s}$ and $k^\mathrm{s} = R_x (1 - m_0^\mathrm{s})$.

\subsection{Sequence-averaged RF saturation approximation}

Following the continuous-wave-equivalent formulation of Teixeira et al.~\cite{Rui2019}, the semisolid-pool saturation rate can be expressed as

\begin{equation}
\langle\overline{W}\rangle = \pi \left(\gamma B_{1,\mathrm{rms}}\right)^2 G(\Delta),
\label{eq:w_cw}
\end{equation}
where $\gamma$ is the gyromagnetic ratio, $B_{1,\mathrm{rms}}$ is the root-mean-square transmit RF amplitude, and $G(\Delta)$ is the absorption lineshape of the semisolid pool at frequency offset $\Delta$.

Teixeira et al.~\cite{Rui2019} evaluated Eq.~\eqref{eq:w_cw} for short-$\mathrm{TR}$ steady-state acquisitions, defining $B_{1,\mathrm{rms}}$ over a single repetition period. To apply the same analytical framework to conventional clinical MRI, we summarized the cumulative RF power of all excitation, refocusing, and inversion pulses by a sequence-averaged $B_{1,\mathrm{rms}}$ computed over the entire acquisition. This extension assumes that the combined RF pulse rate is high relative to the longitudinal relaxation, exchange, and semisolid-pool saturation rates. Under this assumption, the discrete RF pulse train can be approximated by continuous irradiation characterized by the sequence-averaged RF power. Appendix~\ref{sec:appendix_pulse_rate} provides further justification.



Assuming a (i) Gaussian absorption lineshape for the semisolid pool~\cite{Henkelman1993, Ramani2002}, $G(\Delta) = (T_2^\mathrm{s}/\sqrt{2\pi}) \exp[-(2\pi\Delta T_2^\mathrm{s})^2 / 2]$, and (ii) RF irradiation near the water resonance, the lineshape can be evaluated at its peak value, $G(\Delta)\approx G(0)=T_2^\mathrm{s}/\sqrt{2\pi}$. Eq.~\eqref{eq:w_cw} then reduces to

\begin{equation}
\langle\overline{W}\rangle = \sqrt{\pi/2}\,\gamma^2 T_2^\mathrm{s} B_{1,\mathrm{rms}}^2.
\label{eq:w_simplified}
\end{equation}

Appendix~\ref{sec:appendix_lineshape_approx} further discusses the validity of the near-resonance assumption for the clinical acquisitions considered in this work.




\subsection{MT-informed forward signal modeling}
\label{subsec:forward_models}

\begin{figure}[tb]
    \centering
    \includegraphics[width=0.6\linewidth]{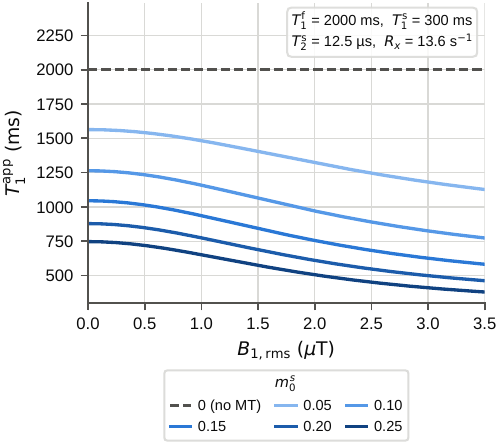}
    \caption{Theoretical behavior of apparent longitudinal relaxation time ($T_1^{\mathrm{app}} = 1/R_1^{\mathrm{app}}$) as a function of sequence-averaged RF amplitude ($B_{1,\mathrm{rms}}$) across varying macromolecular pool fractions ($m_0^\mathrm{s} \in [0, 0.25]$) according to Eq.~\eqref{eq:r1_apparent}. Fixed representative white matter parameters: $T_1^\mathrm{f} = 2000~\mathrm{ms}$, $T_1^\mathrm{s} = 300~\mathrm{ms}$, $T_2^\mathrm{s} = 12.5~\mu\mathrm{s}$, and $R_x = 13.6~\mathrm{s}^{-1}$. The dashed line indicates the free-water longitudinal relaxation time ($T_1^\mathrm{f}$) in the absence of MT ($m_0^\mathrm{s} = 0$).}
    \label{fig:t1app_vs_b1rms}
\end{figure}


Under the continuous-wave-equivalent approximation of Teixeira et al.~\cite{Rui2019}, the two-pool longitudinal exchange reduces to an effective single-pool relaxation process. Expressing the exchange rate constants in terms of $R_x$ and $m_0^\mathrm{s}$ yields the apparent longitudinal relaxation rate:

\begin{equation}
R_1^{\mathrm{app}} = R_1^\mathrm{f} + R_x m_0^\mathrm{s} \left( 1 - \frac{R_x (1 - m_0^\mathrm{s})}{R_1^\mathrm{s} + R_x (1 - m_0^\mathrm{s}) + \langle\overline{W}\rangle} \right),
\label{eq:r1_apparent}
\end{equation}

where $\langle\overline{W}\rangle$ (Eq.~\eqref{eq:w_simplified}) couples sequence-specific RF power ($B_{1,\mathrm{rms}}$) directly to the observed longitudinal relaxation rate. Figure~\ref{fig:t1app_vs_b1rms} illustrates the theoretical dependence of $T_1^{\mathrm{app}} = 1/R_1^{\mathrm{app}}$ on $B_{1,\mathrm{rms}}$ across representative biological macromolecular pool fractions ($m_0^\mathrm{s}$). 

By substituting $R_1^{\mathrm{app}}$ into established Bloch-based signal models \cite{Warntjes2008}, conventional acquisitions can be modeled in closed form while accounting for sequence-specific MT effects. For a given voxel, the expected signal intensity $S$ is expressed as a function of tissue parameters ($\mathrm{PD}$, $R_1^\mathrm{f} = 1/T_1^\mathrm{f}$, $T_2^\mathrm{f}$, $m_0^\mathrm{s}$) and acquisition parameters (repetition time ($\mathrm{TR}$), echo time ($\mathrm{TE}$), inversion time ($\mathrm{TI}$), flip angle ($\mathrm{FA}$), and sequence-averaged $B_{1,\mathrm{rms}}$):

\begin{align}
S_{\mathrm{T1w,GRE}} &= \mathrm{PD} \cdot \sin(\mathrm{FA}) \cdot \left[ \frac{1 - \exp(-R_1^{\mathrm{app}} \mathrm{TR})}{1 - \cos(\mathrm{FA}) \exp(-R_1^{\mathrm{app}} \mathrm{TR})} \right] \cdot \exp(-\mathrm{TE} / T_2^{\mathrm{f}}), \label{eq:siggre} \\[8pt]
S_{\mathrm{T1w,MPRAGE}} &= \left| \mathrm{PD} \cdot \sin(\mathrm{FA}) \cdot \left[ \frac{1 - 2\exp(-R_1^{\mathrm{app}} \mathrm{TI}) + \exp(-R_1^{\mathrm{app}} \mathrm{TR})}{1 + \cos(\mathrm{FA}) \exp(-R_1^{\mathrm{app}} \mathrm{TR})} \right] \right| \cdot \exp(-\mathrm{TE} / T_2^{\mathrm{f}}), \label{eq:sigmprage} \\[8pt]
S_{\mathrm{T2w}} &= \mathrm{PD} \cdot \left[ 1 - \exp(-R_1^{\mathrm{app}} \mathrm{TR}) \right] \cdot \exp(-\mathrm{TE} / T_2^{\mathrm{f}}), \label{eq:sigtse} \\[8pt]
S_{\mathrm{FLAIR}} &= \left| \mathrm{PD} \cdot \left[ 1 - 2\exp(-R_1^{\mathrm{app}} \mathrm{TI}) + \exp(-R_1^{\mathrm{app}} \mathrm{TR}) \right] \right| \cdot \exp(-\mathrm{TE} / T_2^{\mathrm{f}}). \label{eq:sigflair}
\end{align}

Here, $R_1^{\mathrm{app}}$ is computed individually for each sequence via Eq.~\eqref{eq:r1_apparent} using its respective $B_{1,\mathrm{rms}}$, with fixed literature values assumed for the semisolid relaxation and exchange parameters ($R_1^\mathrm{s}$, $R_2^\mathrm{s}$, and $R_x$; detailed in Section~\ref{sec:objective}). For the gradient-echo-based acquisitions, $T_2^\mathrm{f}$ was used as an approximation of $T_2^{*\mathrm{f}}$.

Although the two-pool reduction formally also yields a sequence-dependent apparent equilibrium magnetization ($M_0^{\mathrm{app}}$)~\cite{Rui2019}, we represented its role in the signal models using a single sequence-independent relative $\mathrm{PD}$ map. Under this simplification, sequence-dependent MT effects are modeled only through $R_1^{\mathrm{app}}$. The rationale for this simplification is discussed in Appendix~\ref{sec:appendix_lineshape_approx}.

\section{Methods}

\subsection{Dataset}

\subsubsection{Clinical dataset}
We compiled a multicenter clinical dataset comprising 7,705 non-contrast-enhanced 3~T brain MRI sessions from 4,489 unique subjects from two different sources. The dataset combined (1) retrospectively collected examinations from the University Medical Center Utrecht (UMCU) with (2) examinations from the publicly available MR-RATE dataset acquired at Istanbul Medipol University\footnote{\url{https://huggingface.co/datasets/Forithmus/MR-RATE}}. Each session included a 3D $T_1$-weighted ($T_1\mathrm{w}$) acquisition, an axial 2D $T_2$-weighted ($T_2\mathrm{w}$) acquisition, and a 3D $T_2$-weighted fluid-attenuated inversion recovery (FLAIR) acquisition. Gadolinium-enhanced $T_1\mathrm{w}$ acquisitions were excluded. The two data sources were combined to broaden the distribution of scanner systems, pulse-sequence implementations, and acquisition parameters, including $B_{1,\mathrm{rms}}$. Table~\ref{tab:table_protocol_stacked} summarizes the acquisition characteristics of both datasets.

The UMCU dataset expanded upon the clinical dataset used in our previous work~\cite{vanLune2026} by including additional examinations acquired with a broader range of protocols. The collected data were acquired between 2016 and 2023 and processed in accordance with institutional regulations and ethical guidelines. The resulting dataset comprised 5,312 MRI sessions from 2,192 unique subjects (age range, 6--91 years; mean $\pm$ standard deviation, $51.5 \pm 17.6$ years; 51.1\% male and 48.9\% female). The dataset included patients with a broad range of clinical conditions, including oncological, neurodegenerative, epileptic, and vascular disorders. Examinations were performed on four different Philips 3~T MRI systems: Achieva (33.4\% of sessions), Ingenia (28.4\%), Ingenia CX (20.6\%), and Ingenia Elition X (17.6\%). The $T_1\mathrm{w}$ acquisitions used either a 3D spoiled gradient-echo sequence (Spoiled GRE: 78.5\%) or 3D magnetization-prepared rapid gradient-echo (MPRAGE; 21.5\%). All $T_2\mathrm{w}$ and FLAIR acquisitions used 2D and 3D turbo spin-echo (TSE) sequences, respectively.

A subset of the MR-RATE data was selected based on the availability of the sequence parameters required for the forward signal model: TE, TR, inversion time TI, FA, and root-mean-square RF transmit field strength over the sequence ($B_{1,\mathrm{rms}}$). These criteria yielded 2,393 MRI sessions from 2,297 unique subjects, acquired between 2023 and 2026 (age range, 13--93 years; mean $\pm$ standard deviation, $42.0 \pm 17.7$ years; 43.2\% male and 56.8\% female). All selected examinations were acquired on a Siemens Magnetom Vida 3~T system using 3D MPRAGE for $T_1\mathrm{w}$, 2D TSE for $T_2\mathrm{w}$, and 3D TSE for FLAIR. 

For each dataset, unique subjects were randomly assigned to the training, validation, and test sets using proportions of 80\%, 5\%, and 15\%, respectively. All MRI sessions from the same subject were assigned to the same partition to prevent subject-level data leakage. We then combined the corresponding partitions from both datasets for model training, validation, and testing. The final training set comprised 6,084 sessions from 3,595 subjects, the validation set comprised 384 sessions from 225 subjects, and the test set comprised 1,237 sessions from 669 subjects. Table~\ref{tab:dataset_splits} summarizes the partition sizes. 

MR-RATE provides pathology labels derived automatically from the accompanying radiology reports. Within the MR-RATE test set, 70 sessions from 65 subjects carried the label ``demyelinating disease of the central nervous system'' (mean age, $34.4 \pm 11.0$ years; 35.4\% male and 64.6\% female). Because diagnostic labels were not available for the UMCU dataset, the radiologist in our team (J.W.D.) performed a targeted review and identified four test-set subjects with clinically confirmed MS. This targeted review was not intended to identify all subjects with MS in the UMCU test set. Supplementary Table~\ref{tab:umcu_ms_acquisitions} provides the demographic and acquisition characteristics of these four subjects.

\begin{table}[bt]
\centering
\small
\setlength{\tabcolsep}{3pt}
\renewcommand{\arraystretch}{1.12}
\caption{Acquisition parameters for the UMCU and MR-RATE clinical datasets (training, validation, and test sets combined). Values denote median [min--max]; single values indicate dataset-wide constants. GRE: Spoiled Gradient-Echo; MPRAGE: Magnetization-Prepared Rapid Gradient-Echo; TSE: Turbo Spin Echo; IR: Inversion Recovery; NA: Not Applicable; FA: Flip Angle; ETL: Echo Train Length; BW: Pixel Bandwidth.}
\label{tab:table_protocol_stacked}

\textbf{UMCU (Philips)}\par\smallskip
\begin{tabular*}{\linewidth}{@{\extracolsep{\fill}} l cccc @{}}
\toprule
& \multicolumn{2}{c}{\boldmath$\boldsymbol{T_1\mathrm{w}}$} & \boldmath$\boldsymbol{T_2\mathrm{w}}$ & \textbf{FLAIR} \\
\cmidrule(lr){2-3} \cmidrule(lr){4-4} \cmidrule(lr){5-5}
\textbf{Sequence} & 3D Spoiled GRE & 3D MPRAGE & 2D TSE & 3D TSE (IR) \\
\midrule
\textbf{Sessions ($n$)} & 4,168 & 1,144 & 5,312 & 5,312 \\
\textbf{TR [ms]} & 5.26 [5.03--5.40] & 2500 & 4063 [2779--4918] & 4800 \\
\textbf{TE [ms]} & 2.38 [2.25--2.49] & 3.94 [3.53--4.09] & 80 & 314.5 [275.0--370.3] \\
\textbf{TI [ms]} & NA & 1004 & NA & 1650 \\
\textbf{FA [$^\circ$]} & 10 & 8 & 90 & 90 \\
\textbf{\boldmath $B_{1,\mathrm{rms}}$ [$\mu$T]} & 0.93 [0.92--0.95] & 0.73 [0.73--0.91] & 2.34 [2.26--2.34] & 1.00 [1.00--1.00] \\
\textbf{ETL} & NA & NA & 15 [15--18] & 199 \\
\textbf{BW [Hz/px]} & 539 [539--543] & 190 [189--212] & 195 [177--548] & 1201 [863--1617] \\
\textbf{Matrix (in-plane)} & $480^2$ [$480^2$--$560^2$] & $240^2$ [$224^2$--$288^2$] & $512^2$ [$448^2$--$560^2$] & $240^2$ [$224^2$--$288^2$] \\
\textbf{Slices ($n$)} & 360 [306--480] & 160 [148--240] & 37 [18--46] & 267 [229--333] \\
\bottomrule
\end{tabular*}

\vspace{0.8em}
\textbf{MR-RATE (Siemens)}\par\smallskip
\begin{tabular*}{\linewidth}{@{\extracolsep{\fill}} l ccc @{}}
\toprule
& \boldmath$\boldsymbol{T_1\mathrm{w}}$ & \boldmath$\boldsymbol{T_2\mathrm{w}}$ & \textbf{FLAIR} \\
\cmidrule(lr){2-2} \cmidrule(lr){3-3} \cmidrule(lr){4-4}
\textbf{Sequence} & 3D MPRAGE & 2D TSE & 3D TSE (IR) \\
\midrule
\textbf{Sessions ($n$)} & 2,393 & 2,393 & 2,393 \\
\textbf{TR [ms]} & 2000 [1780--2080] & 5000 [3000--10350] & 7000 [5200--7000] \\
\textbf{TE [ms]} & 2.37 [2.20--4.71] & 95 [89--106] & 388 [386--393] \\
\textbf{TI [ms]} & 920 [920--1050] & NA & 2050 \\
\textbf{FA [$^\circ$]} & 9 [8--9] & 150 [119--150] & 120 \\
\textbf{\boldmath $B_{1,\mathrm{rms}}$ [$\mu$T]} & 0.78 [0.76--0.92] & 1.96 [1.61--2.64] & 1.11 [1.10--1.28] \\
\textbf{ETL} & NA & 17 [13--20] & 220 [218--220] \\
\textbf{BW [Hz/px]} & 248 [199--248] & 220 [216--252] & 751 [651--751] \\
\textbf{Matrix (in-plane)} & $320^2$ [$256^2$--$640^2$] & $512^2$ [$320^2$--$1024^2$] & $512^2$ [$512\times 432$--$568\times 512$] \\
\textbf{Slices ($n$)} & 176 [144--320] & 33 [16--58] & 176 [144--208] \\
\bottomrule
\end{tabular*}
\end{table}

\begin{table}[bt]
\centering
\small
\setlength{\tabcolsep}{4.5pt}
\renewcommand{\arraystretch}{1.15}
\caption{Subject-level partitioning of the UMCU, MR-RATE, and combined clinical datasets. Each source dataset was independently partitioned using an 80/5/15\% training, validation, and test split.}
\label{tab:dataset_splits}
\begin{tabular*}{\linewidth}{@{\extracolsep{\fill}} ll rrrr @{}}
\toprule
\textbf{Dataset} & \textbf{Level} & \textbf{Training (80\%)} & \textbf{Validation (5\%)} & \textbf{Test (15\%)} & \textbf{Total} \\
\midrule
\multirow{2}{*}{UMCU (Philips)} & Sessions & 4,173 & 265 & 874 & 5,312 \\
                                & Subjects & 1,755 & 110 & 327 & 2,192 \\
\addlinespace[0.5em]
\multirow{2}{*}{MR-RATE (Siemens)} & Sessions & 1,911 & 119 & 363 & 2,393 \\
                                   & Subjects & 1,840 & 115 & 342 & 2,297 \\
\midrule
\multirow{2}{*}{\textbf{Combined}} & Sessions & 6,084 & 384 & 1,237 & 7,705 \\
                                   & Subjects & 3,595 & 225 & 669 & 4,489 \\
\bottomrule
\end{tabular*}
\end{table}

\subsubsection{Data preprocessing}
The UMCU MRIs were converted from DICOM to NIfTI format using dcm2niix \cite{Li2016}. We extracted sequence parameters from the DICOM headers. The MR-RATE dataset was already in NIfTI format, and we obtained the corresponding sequence parameters from the accompanying metadata files.

All MRI sessions from the combined dataset underwent a series of preprocessing steps. First, all $T_2\mathrm{w}$ acquisitions were resampled to an in-plane resolution of $1 \times 1\text{ mm}$ while preserving their native slice thickness. Across the combined dataset, the native $T_2\mathrm{w}$ slice thickness was either 3.0 or 4.0 mm. Second, the 3D $T_1\mathrm{w}$ and FLAIR images were rigidly registered and resampled onto the voxel grid of the corresponding resampled 2D $T_2\mathrm{w}$ acquisition using ANTsPy~\cite{Avants2009}. Third, a brain mask was generated from the $T_1\mathrm{w}$ image using HD-BET \cite{Isensee2019}, and all images were skull stripped using the resulting mask. Fourth, N4 bias-field correction~\cite{Tustison2010} was applied to the skull-stripped $T_1\mathrm{w}$, $T_2\mathrm{w}$, and FLAIR images. Fifth, axial slices were center-cropped or zero-padded to a matrix size of $224 \times 224$ voxels. Finally, superior and inferior slices with low signal-to-noise ratio were excluded to reduce artifacts. Before training and inference of our proposed framework, each input image was independently normalized by dividing all voxel intensities by the histogram mode of the corresponding whole-brain volume.

To obtain tissue masks, we applied SynthSeg~\cite{Billot2023} to the preprocessed $T_1\mathrm{w}$ images from each MRI session. White matter (WM) and gray matter (GM) masks were extracted from the resulting segmentations. To reduce partial-volume contamination from cerebrospinal fluid (CSF), we excluded WM and GM voxels located within $2.0~\mathrm{mm}$ of CSF.


For the demyelinating lesion analyses, we segmented WM lesions using LST-AI, a deep learning model developed and trained for MS lesion segmentation on paired $T_1\mathrm{w}$ and FLAIR images~\cite{Wiltgen2024}. Segmentation was performed for the four manually identified UMCU subjects with MS and the 70 MR-RATE MRI sessions with the demyelinating disease label. LST-AI was applied to the native-resolution, full-head $T_1\mathrm{w}$ and FLAIR acquisitions selected from each MRI session before any preprocessing steps. The resulting lesion masks were aligned with the preprocessed images using the same rigid transformation and resampling procedure applied during preprocessing. Lesion voxels were excluded from the WM masks to define normal-appearing white matter (NAWM) and gray matter (NAGM) for these MRI sessions.

\subsection{Model}
We extended our previously proposed physics-guided self-supervised framework for retrospective qMRI mapping from conventional clinical scans~\cite{vanLune2026} by incorporating the MT-informed apparent longitudinal relaxation model into the differentiable forward signal model (Eqs.~\eqref{eq:r1_apparent}--\eqref{eq:sigflair}). This extension enables the multi-contrast convolutional neural network (CNN) to estimate voxel-wise maps of the free-pool longitudinal relaxation time ($T_1^\mathrm{f}$), free-pool transverse relaxation time ($T_2^\mathrm{f}$), $\mathrm{PD}$, and macromolecular pool fraction ($m_0^\mathrm{s}$; MPF) directly from routine $T_1\mathrm{w}$, $T_2\mathrm{w}$, and FLAIR acquisitions. Training remains self-supervised and does not require reference quantitative maps, which are not available.

\subsubsection{Network architecture}
The network retains the three-level 2D U-Net architecture described in~\cite{vanLune2026}. A shared-weight encoder processes each input contrast of an MRI session separately, and attention-based contrast-fusion modules combine the resulting feature representations at each resolution level. The fused features are passed to a common decoder with skip connections from the corresponding encoder levels. The network operates on 2D axial slices of $224 \times 224$ voxels.

We modified the original architecture in two respects. First, we expanded the output layer from three to four channels to generate $T_1^\mathrm{f}$, $T_2^\mathrm{f}$, $\mathrm{PD}$, and $m_0^\mathrm{s}$ maps. A Sigmoid activation function constrained the outputs to $[0,1]$, after which they were linearly scaled to physical ranges of $[0,5]$~s for $T_1^\mathrm{f}$, $[0,3]$~s for $T_2^\mathrm{f}$, $[0,1]$ arbitrary units for $\mathrm{PD}$, and $[0,0.4]$ for $m_0^\mathrm{s}$. We otherwise kept the decoder and attention-based contrast-fusion modules unchanged.

Second, we replaced the standard instance normalization in the encoder and bottleneck with acquisition-conditioned adaptive instance normalization (AdaIN)~\cite{dumoulin2017learned,huang2017adain,karras2019style}. 
Conditioning neural networks on sequence parameters has previously been shown to mitigate protocol-dependent variability across heterogeneous protocols~\cite{Chen2026,Qiu2024ISMRM,Borges2023}. Here, AdaIN enables the shared encoder to dynamically modulate its intermediate feature representations based on contrast type and sequence timing parameters.


To condition the network on the acquisition settings, each input contrast $m \in \{T_1\mathrm{w}, T_2\mathrm{w}, \mathrm{FLAIR}\}$ is paired with a nine-dimensional acquisition parameter vector $\mathbf{c}_m \in \mathbb{R}^9$:

\begin{equation} \mathbf{c}_m = \left[ c_{T_1\mathrm{w}},\, c_{T_2\mathrm{w}},\, c_{\mathrm{FLAIR}},\, \ln(\mathrm{TR}/1\,\mathrm{ms}),\, \ln(\mathrm{TE}/1\,\mathrm{ms}),\, \sin(\mathrm{FA}),\, B_{1,\mathrm{rms}},\, c_{\mathrm{TI}},\, \ln(\mathrm{TI}/1\,\mathrm{ms}) \right]^\top.
\label{eq:conditioning_vector} 
\end{equation}

The first three terms are binary indicator variables ($c_{T_1\mathrm{w}}, c_{T_2\mathrm{w}}, c_{\mathrm{FLAIR}} \in \{0, 1\}$), with exactly one set to 1 to identify the contrast type. The continuous timing parameters $\mathrm{TR}$, $\mathrm{TE}$, and $\mathrm{TI}$ are log-transformed to compress their numerical ranges. The flip angle $\mathrm{FA}$ is transformed via $\sin(\mathrm{FA})$ and $B_{1,\mathrm{rms}}$ is the sequence-averaged root-mean-square RF transmit amplitude in $\mu\mathrm{T}$. The binary indicator $c_{\mathrm{TI}} \in \{0, 1\}$ denotes whether an inversion preparation pulse was applied ($c_{\mathrm{TI}} = 1$ for MPRAGE and FLAIR; $c_{\mathrm{TI}} = 0$ for non-inversion GRE and TSE acquisitions). When $c_{\mathrm{TI}} = 0$, the corresponding logarithmic inversion time entry is set to zero.

A two-layer multilayer perceptron (MLP) with exponential linear unit (ELU) activations~\cite{Clevert2015} projects $\mathbf{c}_m$ to a 128-dimensional acquisition embedding $\mathbf{w}_m$. For an intermediate feature map $\mathbf{x} \in \mathbb{R}^{C \times H \times W}$ with channel-wise mean $\mu_c(\mathbf{x})$ and standard deviation $\sigma_c(\mathbf{x})$, AdaIN normalizes each channel and applies sequence-specific affine modulation:

\begin{equation}
\mathrm{AdaIN}(\mathbf{x},\mathbf{w}_m)_c = \gamma_c(\mathbf{w}_m) \left[ \frac{\mathbf{x}_c - \mu_c(\mathbf{x})}{\sigma_c(\mathbf{x})} \right] + \beta_c(\mathbf{w}_m),
\label{eq:adain}
\end{equation}

where the channel-wise scale and shift vectors are generated via learned linear projections:

\begin{equation}
\boldsymbol{\gamma}(\mathbf{w}_m) = \mathbf{W}_\gamma \mathbf{w}_m + \mathbf{b}_\gamma + \mathbf{1}, \qquad \boldsymbol{\beta}(\mathbf{w}_m) = \mathbf{W}_\beta \mathbf{w}_m + \mathbf{b}_\beta.
\label{eq:adain_projections}
\end{equation}

Both the projection weights and biases are initialized to zero. This ensures that each AdaIN block acts as standard instance normalization at the onset of training and progressively learns acquisition-dependent feature modulation. The conditioning MLP and AdaIN layers added approximately 0.50M trainable parameters, corresponding to 5.6\% of the 9.06M trainable parameters of the full network.

\subsubsection{Self-supervised physics-guided objective}
\label{sec:objective}
The framework was trained using a physics-guided self-supervised objective that enforced consistency between the acquired input clinical scans and synthetic images generated by the forward signal model. For each MRI session, the network predicted voxel-wise maps of $\mathrm{PD}$, $T_1^\mathrm{f}$, $T_2^\mathrm{f}$, and $m_0^\mathrm{s}$. The MT-informed forward signal models described in Section~\ref{subsec:forward_models} then combined these maps with the acquisition parameters of each input scan, including $\mathrm{TR}$, $\mathrm{TE}$, $\mathrm{TI}$, $\mathrm{FA}$, and $B_{1,\mathrm{rms}}$, to synthesize images $\hat{S}_m$ for contrasts $m \in \{T_1\mathrm{w}, T_2\mathrm{w}, \mathrm{FLAIR}\}$ using Eqs.~\eqref{eq:siggre}--\eqref{eq:sigflair}. Semisolid-pool and exchange parameters that were not estimated by the network were fixed to values from recent qMT literature for WM at 3~T: $\gamma = 2.675 \times 10^8\,\text{rad}\,\text{s}^{-1}\text{T}^{-1}$, $R_x = 13.6\,\text{s}^{-1}$, $T_1^\mathrm{s} = 300\,\text{ms}$, and $T_2^\mathrm{s} = 12.5\,\mu\text{s}$~\cite{Asslnder2024a, Asslnder2025var, vanGelderen2016}.

For the TSE-based $T_2\mathrm{w}$ and FLAIR acquisitions, the forward signal model required an additional correction for effective $T_2$ weighting. Specifically, the analytical Bloch-based equations are less accurate for TSE acquisitions because these sequences use long echo trains with variable refocusing flip angles, so the effective $T_2$ weighting differs from that predicted by the nominal echo time. To account for this difference, we calculated TE correction factors following Busse et al.~\cite{Busse2006} and applied them to the nominal TE values in the forward signal model. The factors were 0.90 and 0.42 for the Philips $T_2\mathrm{w}$ and FLAIR acquisitions, respectively, and 0.86--0.96 and approximately 0.51 for the corresponding Siemens acquisitions. Supplementary Section~\ref{sec:appendix_tse_echo_time} details the calculations.

Conventional MRI signal intensities are expressed in arbitrary, non-standardized units and therefore contain an unknown scan-dependent global scaling factor. Before computing the training loss, each synthesized image $\hat{S}_m$ was globally rescaled to the corresponding normalized input image $S_m$ using the calibration procedure described in our previous work~\cite{vanLune2026}. Because the whole-brain histogram mode used for input normalization corresponds to WM intensity, the scaling factor was constrained to within 20\% of a reference value derived from representative WM tissue parameters: $T_1^\mathrm{f} = 2000\,\text{ms}$, $T_2^\mathrm{f} = 70\,\text{ms}$, $\mathrm{PD} = 0.70$, and $m_0^\mathrm{s} = 0.21$~\cite{Asslnder2024a, Asslnder2025var, vanGelderen2016, Bojorquez2017}.

In the Bloch-based forward signal models, $T_1^\mathrm{f}$ and $m_0^\mathrm{s}$ both affect the synthesized signal intensity through $R_1^\mathrm{app}$ (Eq.~\eqref{eq:r1_apparent}), resulting in an identifiability problem. For a fixed $B_{1,\mathrm{rms}}$, multiple combinations of $T_1^\mathrm{f}$ and $m_0^\mathrm{s}$ can therefore produce identical image contrast. Variation in $B_{1,\mathrm{rms}}$ across the dataset helps constrain this ambiguity. To further separate the two parameters, we imposed a prior on $T_1^\mathrm{f}$ within WM only, with $\mu_{T_1}=2000\,\text{ms}$ and $\sigma_{T_1}=400\,\text{ms}$. This prior mean reflects the free-water relaxation time at 3~T reported in recent qMT literature~\cite{Asslnder2024a, vanGelderen2016}, which is substantially longer than the apparent tissue $T_1$ ($\sim 700\text{--}1100\,\text{ms}$~\cite{Bojorquez2017}). No direct prior or regularization term was imposed on $m_0^\mathrm{s}$.

As in our previous work~\cite{vanLune2026}, a soft lower bound was imposed on $\mathrm{PD}$ to constrain the solution space of the ill-posed inverse problem. This constraint was implemented as a hinge penalty with a threshold $\tau = 0.60$. The total training loss combined a data-consistency image-reconstruction term with the $\mathrm{PD}$ constraint and the white-matter $T_1^\mathrm{f}$ prior regularization terms:

\begin{equation}
\mathcal{L} = \sum_{m} \left\lVert \tilde{S}_m - S_m \right\rVert_1 +\lambda_{\mathrm{PD\text{-}constraint}} \left\langle \max\left(0,\, \tau - \mathrm{PD} \right) \right\rangle + \lambda_{T_1\text{-}\mathrm{prior}} \left\langle \frac{\left(T_1^\mathrm{f} - \mu_{T_1}\right)^2}{2\sigma_{T_1}^2} \right\rangle_{\mathrm{WM}},
\label{eq:loss}
\end{equation}

where $\tilde{S}_m$ and $S_m$ denote the rescaled synthesized and normalized acquired images for contrast $m \in \{T_1\mathrm{w}, T_2\mathrm{w}, \mathrm{FLAIR}\}$, respectively. The angle brackets $\langle \cdot \rangle$ represent spatial voxel averaging, with $\langle \cdot \rangle_{\mathrm{WM}}$ restricted to a white matter mask. The hyperparameters $\lambda_{\mathrm{PD\text{-}constraint}}$ and $\lambda_{T_1\text{-}\mathrm{prior}}$ control the relative contributions of the two regularization terms. 

\subsubsection{Training procedure and implementation details}
The network was trained on the full combined dataset for 20 epochs using the Adam optimizer \citep{Kingma2014} with momentum parameters $\beta_1 = 0.9$ and $\beta_2 = 0.999$. Network weights were initialized using Kaiming initialization \citep{He2015}. Each training batch consisted of slices randomly sampled from the entire training dataset. To identify the optimal model configuration, hyperparameter optimization was performed using a grid search. Table~\ref{tab:hyperparams} in the Supplementary Materials summarizes the search space and explored values. Furthermore, the necessity of the two regularization terms ($\text{PD-constraint}$ and $T_1\text{-}\mathrm{prior}$) was evaluated through an ablation study, detailed in Supplementary Materials~\ref{sec:appendix_ablation_study}. Model selection was based on minimizing validation loss while ensuring quantitative outputs matched physiologically expected ranges for human brain qMT literature values in WM and GM \cite{Bojorquez2017, Asslnder2024a, vanGelderen2016}.

The final model was trained using a learning rate of $1\times10^{-5}$, a batch size of 16 slices, and regularization weights $\lambda_{T_1\text{-}\mathrm{prior}}=0.0025$ and $\lambda_{\mathrm{PD\text{-}constraint}}=0.5$. We developed the implementation in Python 3.11.9 using PyTorch 2.1.2 \citep{Paszke2019}, with data loading, preprocessing, and slice sampling handled by TorchIO 0.20.6 \citep{Prez-Garca2021}. Training was performed on a single NVIDIA RTX 4500 Ada Generation GPU and took approximately 12 hours. The complete PyTorch implementation and pretrained network weights are available on GitHub: \url{https://github.com/JelmervanL/MPF-mapping-from-conventional-MRI}.

\subsection{Evaluation}
\label{subsec:evaluation}

We evaluated the biological plausibility and protocol robustness of the generated MPF ($m_0^\mathrm{s}$) maps, their sensitivity to demyelinating pathology, their cross-protocol reproducibility, and the contribution of acquisition-conditioned AdaIN.

\subsubsection{Macromolecular pool fraction in white matter and gray matter}
Estimated MPF values were evaluated in WM and GM across the full clinical test set ($n = 1{,}237$). To assess robustness against scanner and protocol variation, sessions were stratified into three vendor-by-protocol groups based on the $T_1\mathrm{w}$ acquisition: Philips Spoiled GRE, Philips MPRAGE, and Siemens MPRAGE. Stratification was based on the $T_1\mathrm{w}$ sequence because it exhibited the largest protocol variation in the dataset, including differences in pulse-sequence type. In contrast, $T_2\mathrm{w}$ and FLAIR acquisitions consistently employed turbo spin-echo sequences across the entire dataset, differing primarily in timing parameters ($\mathrm{TR}$, $\mathrm{TE}$, and $\mathrm{TI}$). We quantified protocol stability using the between-group coefficient of variation (CV) of tissue-specific mean MPF values. The CV was calculated as the standard deviation of the three group means, each obtained by averaging the per-session means within that group, divided by the unweighted mean of the three group means.

\subsubsection{Demyelinating pathology}
Sensitivity to demyelinating pathology was evaluated using the 70 MR-RATE test sessions with report-derived demyelinating-disease labels and the four UMCU test sessions with confirmed MS. Mean MPF values within segmented WM lesions were compared with NAWM and NAGM using relative percentage differences. 

Furthermore, because $T_1^\mathrm{f}$ and $m_0^\mathrm{s}$ jointly determine $R_1^\mathrm{app}$ in the forward signal model (Eq.~\eqref{eq:r1_apparent}), we also assessed whether their estimates covaried across individual lesions. Following the approach of \citet{Asslnder2024a}, individual lesions were defined as distinct three-dimensional connected components of the WM lesion masks. Components containing fewer than 10 voxels were excluded from the analysis. Each lesion was represented by its median $m_0^\mathrm{s}$ and free-pool longitudinal relaxation rate ($R_1^\mathrm{f}=1/T_1^\mathrm{f}$). Principal component analysis (PCA) of these paired lesion-level estimates quantified the dominant direction of joint variation and the remaining variation orthogonal to this direction.

\subsubsection{Prospective volunteer reproducibility}
As a technical demonstration, prospective cross-protocol reproducibility of the generated MPF maps was evaluated in a 26-year-old male volunteer scanned on a 3~T Philips Ingenia system with written informed consent and institutional review board approval. The volunteer was considered healthy based on the absence of visible lesions on the acquired images. Two clinical protocols, each comprising a separate set of $T_1\mathrm{w}$, $T_2\mathrm{w}$, and FLAIR acquisitions, were performed consecutively within the same scanning session. The protocols differed primarily in the $T_1\mathrm{w}$ sequence, which used Spoiled GRE in one protocol and MPRAGE in the other; detailed acquisition characteristics are provided in Supplementary Table~\ref{tab:prospective_acquisitions_info}. The prospective data were preprocessed using the same pipeline as the clinical dataset. Voxel-wise agreement between the generated MPF maps was assessed across brain parenchyma (WM + GM) using the mean absolute percentage difference, Pearson correlation coefficient ($r$), and concordance correlation coefficient (CCC).

\subsubsection{Impact of acquisition-conditioned AdaIN}
Finally, the contribution of acquisition-conditioned AdaIN was assessed through an ablation experiment in which all AdaIN layers were replaced with standard instance normalization, while all other model and training settings were kept unchanged. Performance was compared using test-set L1 reconstruction error, the between-group CV of tissue MPF estimates across vendor-by-protocol groups, and cross-protocol agreement in the prospective volunteer.

\section{Results}

\subsection{Macromolecular pool fraction in white matter and gray matter}

Across the clinical test set ($n = 1{,}237$), estimated macromolecular pool fraction ($m_0^\mathrm{s}$; MPF) values were consistently higher in WM ($0.182 \pm 0.007$) than in GM ($0.144 \pm 0.009$). Estimates varied little across vendor-by-protocol groups, with between-group CVs of 2.65\% for WM and 5.63\% for GM and substantial overlap of the corresponding within-tissue distributions (Table~\ref{tab:reproducibility_scanner_sequence}; Figure~\ref{fig:distributions_m0s}). Results for all estimated quantitative parameters are provided in Supplementary Table~\ref{tab:appendix_reproducibility_scanner_sequence_all}.

\begin{table}[bt]
\centering
\small
\caption{Mean estimated macromolecular pool fraction ($m_0^\mathrm{s}$; MPF) in white matter (WM) and gray matter (GM) across the full clinical test set and the three vendor-by-protocol groups. Values represent mean $\pm$ standard deviation across MRI sessions. }
\label{tab:reproducibility_scanner_sequence}
\begin{tabular*}{\linewidth}{@{\extracolsep{\fill}} l c ccc@{}}
\toprule
& & \multicolumn{2}{c}{\textbf{Philips (UMCU)}} & \textbf{Siemens (MR-RATE)}  \\
\cmidrule(lr){3-4} \cmidrule(lr){5-5}
\textbf{Tissue} & \textbf{Total} & \textbf{Spoiled GRE} & \textbf{MPRAGE} & \textbf{MPRAGE} \\
& ($n = 1,237$) & ($n = 653$) & ($n = 221$) & ($n = 363$)  \\
\midrule
WM & $0.182 \pm 0.007$ & $0.186 \pm 0.006$ & $0.179 \pm 0.004$ & $0.177 \pm 0.006$ \\
GM & $0.144 \pm 0.009$ & $0.149 \pm 0.005$ & $0.147 \pm 0.005$ & $0.134 \pm 0.006$ \\
\bottomrule
\end{tabular*}
\end{table}

\begin{figure}[tb]
    \centering
    \includegraphics[width=0.6\linewidth]{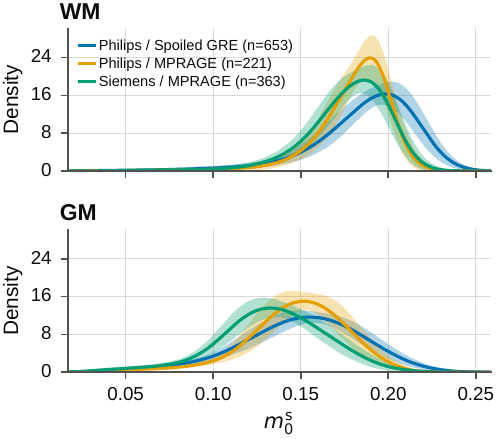}
    \caption{Distribution of estimated macromolecular pool fraction ($m_0^\mathrm{s}$; MPF) values in white matter (WM) and gray matter (GM) for the three vendor-by-protocol groups in the clinical test set. Solid lines show the mean density across MRI sessions and shaded regions indicate one standard deviation.}
\label{fig:distributions_m0s}
\end{figure}

\subsection{Demyelinating pathology}

\begin{figure}[tb]
    \centering
    \includegraphics[width=\linewidth]{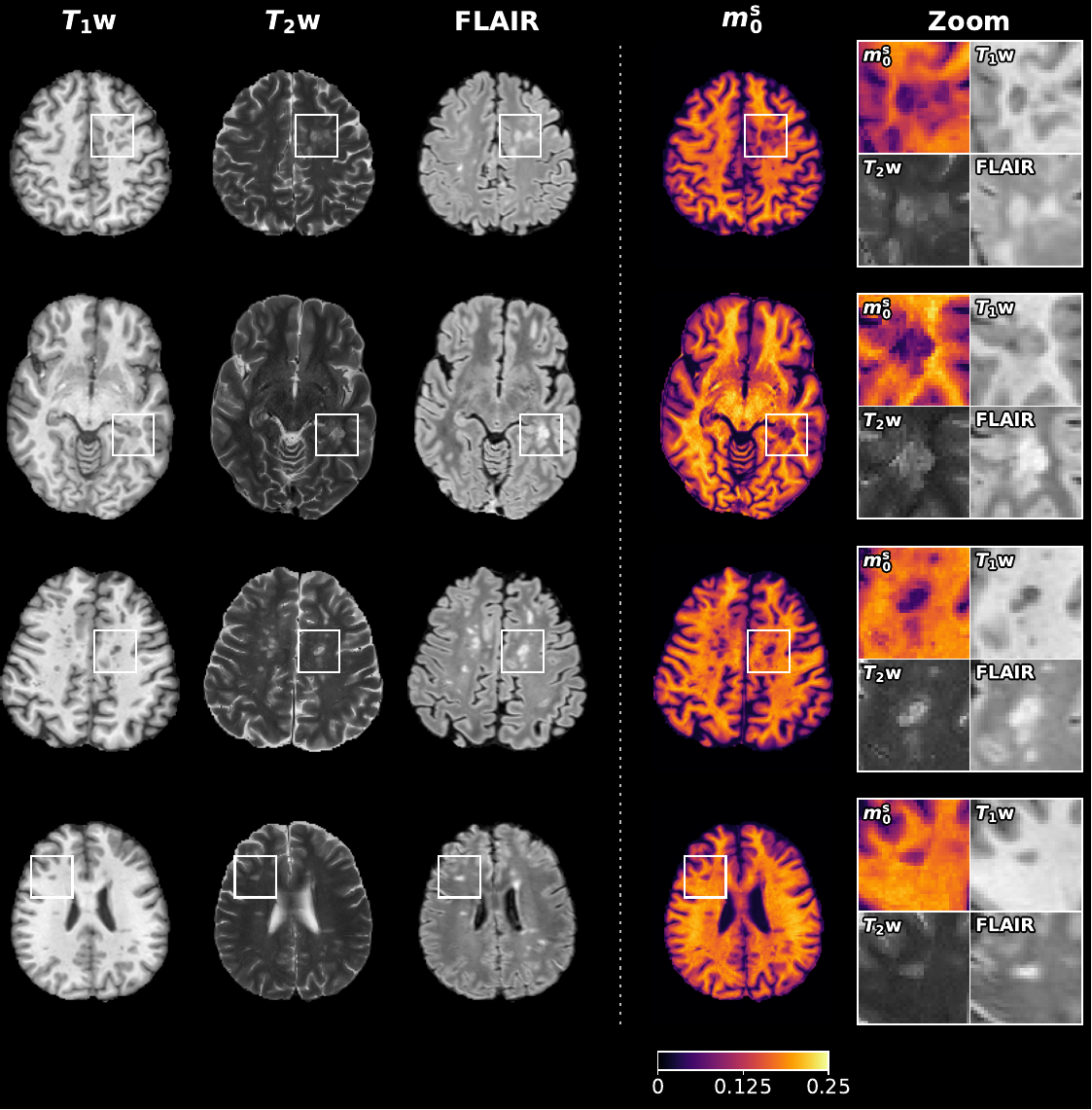}
    \caption{Representative MPF maps from the MR-RATE test set (Siemens MPRAGE). Each of the four rows shows a distinct, randomly selected subject with a report-derived demyelinating-disease diagnosis. Columns show the input $T_1$-weighted, $T_2$-weighted, and FLAIR images together with the generated $m_0^\mathrm{s}$ map. White boxes indicate regions enlarged in the zoom panels.}
\label{fig:montage_MRRATE_ms}
\end{figure}

\begin{figure}[tb]
    \centering
    \includegraphics[width=\linewidth]{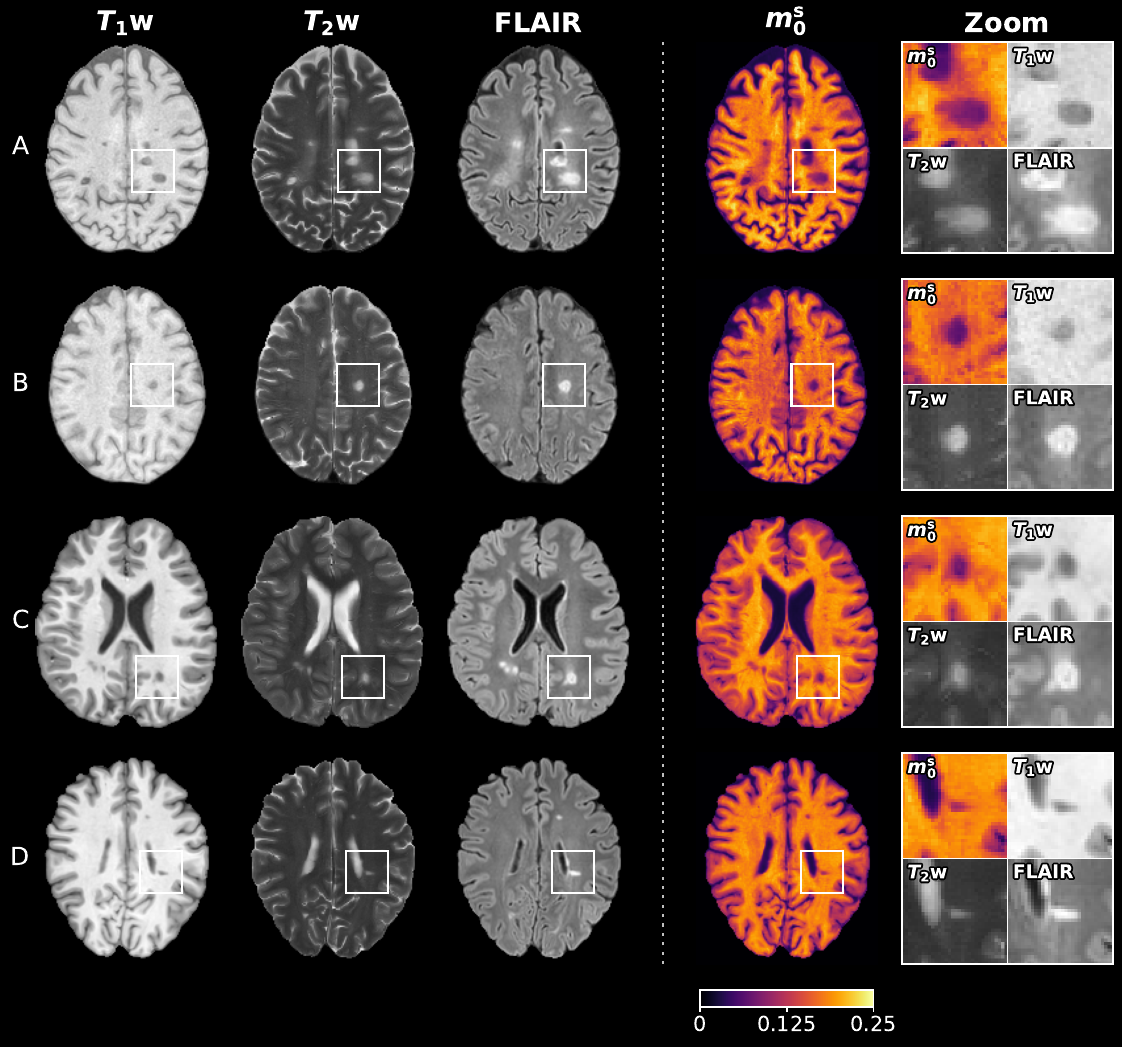}
    \caption{Representative MPF maps from the four subjects with MS identified in the UMCU test set. Rows A and B correspond to the Philips Spoiled GRE vendor-by-protocol group, whereas rows C and D correspond to the Philips MPRAGE vendor-by-protocol group. Columns show the input $T_1$-weighted, $T_2$-weighted, and FLAIR images together with the generated $m_0^\mathrm{s}$ map. White boxes indicate regions enlarged in the zoom panels.}
\label{fig:montage_UMCU_ms}
\end{figure}

In the 70 MR-RATE sessions with a report-derived demyelinating-disease label, mean MPF was $0.113 \pm 0.015$ in segmented WM lesions. Lesion MPF was 35.6\% lower than in NAWM ($0.176 \pm 0.005$) and 15.2\% lower than in NAGM ($0.133 \pm 0.005$). The corresponding results within lesions for $T_1^\mathrm{f}$, $T_2^\mathrm{f}$, and $\mathrm{PD}$ are also provided in Supplementary Table~\ref{tab:appendix_mrrate_lesion_quantification}. Comparable MPF reductions were observed in the four UMCU subjects with MS (Supplementary Table~\ref{tab:appendix_lesion_quantification}). Representative examples from both subsets show that FLAIR-hyperintense lesions corresponded to reductions in MPF relative to the surrounding tissue across scanner vendors and $T_1\mathrm{w}$ acquisition protocols (Figures~\ref{fig:montage_MRRATE_ms} and~\ref{fig:montage_UMCU_ms}).

\begin{figure}[tb]
    \centering
    \includegraphics[width=0.6\linewidth]{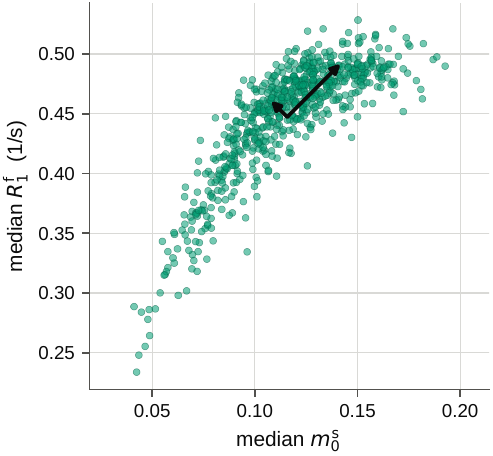}
    \caption{Relationship between lesion-wise free-pool longitudinal relaxation rate ($R_1^\mathrm{f}$) and macromolecular pool fraction ($m_0^\mathrm{s}$; MPF) in the MR-RATE demyelinating-disease subset ($n=70$ MRI sessions, 673 unique lesions). Each point represents the median $R_1^\mathrm{f}$ and $m_0^\mathrm{s}$ value within a single demyelinating lesion. Black arrows indicate the principal-component eigenvectors and are scaled to one standard deviation along each component.}
\label{fig:pca_scatter_r1_m0s}
\end{figure}

To investigate the relationship between free-pool longitudinal relaxation and MPF within demyelinating lesions, we performed PCA on lesion-wise $R_1^\mathrm{f}$ and MPF estimates from the MR-RATE subset (Figure~\ref{fig:pca_scatter_r1_m0s}). The first and second principal components explained 91\% and 9\% of the total variance, respectively.

\subsection{Prospective volunteer reproducibility}

In the controlled prospective cross-protocol evaluation, the generated $m_0^\mathrm{s}$ maps showed similar tissue values across the two acquisition protocols. For the Spoiled GRE protocol, the mean $m_0^\mathrm{s}$ was $0.171 \pm 0.026$ in WM and $0.135 \pm 0.027$ in GM. The corresponding values for the MPRAGE protocol were $0.171 \pm 0.021$ and $0.134 \pm 0.023$, respectively. 

\begin{figure}[tb]
    \centering
    \includegraphics[width=\linewidth]{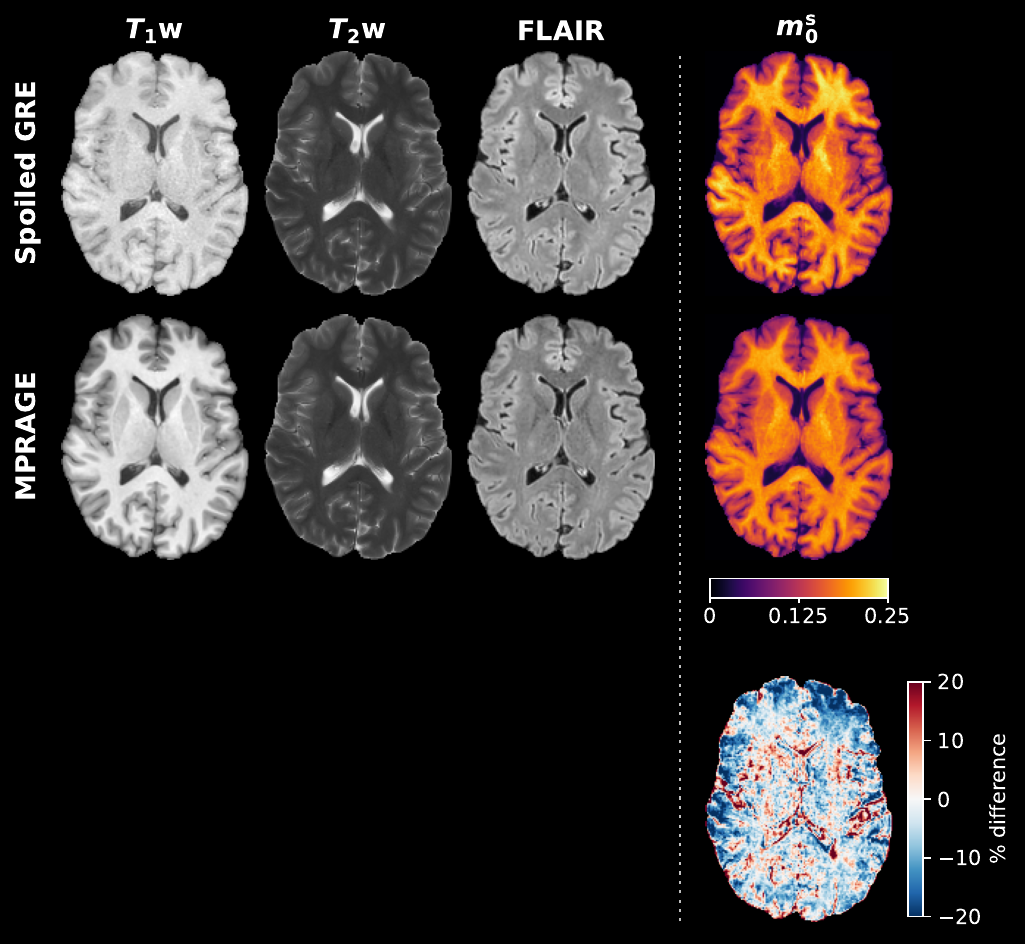}
    \caption{Prospective within-subject reproducibility in a healthy volunteer scanned with two different clinical protocols. Top row: estimated macromolecular pool fraction ($m_0^\mathrm{s}$; MPF) map from the Spoiled GRE protocol. Middle row: corresponding MPF map from the MPRAGE protocol. Bottom row: voxel-wise percentage difference maps between the two generated MPF maps.}
    \label{fig:volunteer_maps}
\end{figure}

\begin{figure}[tb]
    \centering
    \includegraphics[width=0.6\linewidth]{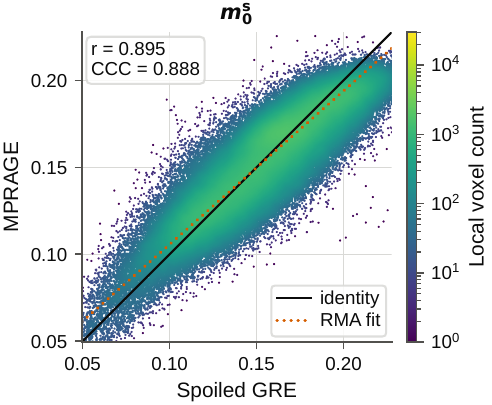}
    \caption{Voxel-wise agreement between macromolecular pool fraction ($m_0^\mathrm{s}$; MPF) maps
    estimated from the Spoiled GRE and MPRAGE protocols in a healthy volunteer. Each point represents a
    brain parenchymal voxel, colored by local voxel density (log scale). The solid line denotes
    identity and the dotted line the reduced major axis (RMA) fit. Pearson's $r$ and the concordance
    correlation coefficient (CCC) are shown in the top left.}
    \label{fig:volunteer_scatter}
\end{figure}

Figure~\ref{fig:volunteer_maps} shows representative MPF maps from both protocols. The voxel-wise percentage-difference map showed only limited regional variation between protocols. Across brain parenchyma (WM+GM), the mean absolute percentage difference between the two generated MPF maps was 7.7\%. The strong agreement between protocols was also evident in the voxel-wise density scatter plot (Figure~\ref{fig:volunteer_scatter}), yielding a Pearson correlation coefficient of $r = 0.895$ and a CCC of $0.888$. Similar reproducibility analyses for the generated $T_1^\mathrm{f}$, $T_2^\mathrm{f}$, and $\mathrm{PD}$ maps are presented in the Supplementary Figures~\ref{fig:appendix_volunteer_slices_full} and~\ref{fig:appendix_volunteer_scatter_full}.

\subsection{Impact of acquisition-conditioned AdaIN}

Replacing acquisition-conditioned AdaIN with standard instance normalization increased the between-group CV of estimated $m_0^\mathrm{s}$ across the three vendor-by-protocol groups. In WM, the CV increased from 2.65\% with AdaIN to 10.27\% without AdaIN; in GM, it increased from 5.63\% to 10.37\%, respectively (additional results are provided in Supplementary Table~\ref{tab:appendix_adain_tissue_comparison}). The mean total L1 reconstruction error across the clinical test set ($n = 1{,}237$ sessions) increased by 22.3\%, from $0.0466 \pm 0.0177$ with AdaIN to $0.0570 \pm 0.0171$ without AdaIN ($p < 0.001$, Wilcoxon signed-rank test).

In the prospective volunteer experiment, the cross-protocol mean absolute percentage difference between the estimated $m_0^\mathrm{s}$ maps increased from 7.7\% with AdaIN to 15.3\% without AdaIN. The voxel-wise CCC decreased from $0.888$ to $0.672$. The corresponding cross-protocol scatter plots are shown in Supplementary Figure~\ref{fig:appendix_volunteer_scatter_adain}.

\section{Discussion}
\label{sec:discussion}

In this work, we introduced a physics-guided, self-supervised deep learning framework for retrospectively estimating MPF from conventional $T_1\mathrm{w}$, $T_2\mathrm{w}$, and FLAIR brain MRI without paired qMT data. The framework extends retrospective quantitative MRI mapping by incorporating an analytical two-pool MT model that relates apparent longitudinal relaxation to sequence-averaged RF power, together with acquisition-conditioned AdaIN to account for differences among acquisition protocols. We trained and evaluated the framework using a large multicenter, multivendor clinical dataset and examined tissue-level MPF estimates, sensitivity to demyelinating pathology, vendor/protocol robustness, and the contribution of AdaIN acquisition conditioning.


We first assessed whether the retrospectively estimated MPF maps showed expected values for WM and GM. Across the clinical test set, estimated MPF was consistently higher in WM ($0.182 \pm 0.007$) than in GM ($0.144 \pm 0.009$), in agreement with the higher macromolecular and myelin content of WM~\cite{Koenig1990, Kisel2022}. These tissue mean values fell within the ranges reported by qMT studies that model distinct longitudinal relaxation rates for the free and semisolid pools ($m_0^\mathrm{s} \in [0.17, 0.27]$ in WM and $m_0^\mathrm{s} \in [0.08, 0.18]$ across cortical and subcortical GM)~\cite{Helms2009, vanGelderen2016, vanGelderen2017, Asslnder2024a, Asslnder2025var}. 
However, absolute MPF values should be compared cautiously across studies, as they depend on the fitting strategy and the assumptions and constraints of the two-pool model. In particular, our estimates were higher than values commonly obtained using models that use the constraint $T_1^\mathrm{s}=T_1^\mathrm{f}$~\cite{Stanisz2005, Yarnykh2012, Yarnykh2016}. Previous studies have shown that this constraint can substantially underestimate the semisolid pool fraction~\cite{Helms2009, vanGelderen2016, Asslnder2024a}. In contrast, our model fixes $T_1^\mathrm{s}$ to the substantially shorter value of $300~\mathrm{ms}$~\cite{Helms2009, vanGelderen2016, Asslnder2024a, Asslnder2025var} and estimates $T_1^\mathrm{f}$ voxel-wise. This difference in model assumptions likely contributes to the higher MPF estimates obtained with our framework. 

Despite the tissue means falling within reported ranges, the contrast between WM and GM was smaller than expected. The GM-to-WM MPF ratio was approximately $0.79$, whereas prospective qMT studies using both constrained and unconstrained two-pool models have generally reported ratios closer to $0.5$~\cite{Stanisz2005, Yarnykh2012, Asslnder2024a}. Our generated maps preserved the expected ordering of WM and GM values but showed reduced MPF contrast between these tissues. This reduced tissue contrast may stem from model simplifications, globally fixed semisolid-pool and exchange parameters, the WM-specific $T_1^\mathrm{f}$ prior, and partial-volume effects, as discussed further in Section~\ref{sec:discussion_limitations}.

In segmented lesions, MPF was 35.6\% lower than in NAWM in the MR-RATE demyelinating-disease subset, consistent with reductions of 23.5--34.2\% in the four UMCU subjects with confirmed MS. These decreases align with prospective quantitative MT literature reporting substantial macromolecular pool reductions in MS lesions~\cite{Asslnder2024a, Kisel2022}, where relative decreases typically range from 30\% to 45\% compared to NAWM~\cite{Davies2004, Cercignani2009, Levesque2010, Yarnykh2015}. This agreement supports the sensitivity of the generated MPF maps to tissue alterations associated with demyelinating pathology.
To assess the extent to which lesion-wise MPF and $R_1^\mathrm{f}$ estimates covaried, we performed PCA on their paired values. The first principal component explained 91\% of their joint variance (Figure~\ref{fig:pca_scatter_r1_m0s}). This proportion was similar to the 93\% reported by \citet{Asslnder2024a} for dedicated prospective qMT under the constraint of equal longitudinal relaxation times for the free and semisolid pools ($T_1^\mathrm{s}=T_1^\mathrm{f}$). Thus, our retrospectively estimated quantitative maps showed a similar degree of joint variation between lesion-wise $R_1^\mathrm{f}$ and MPF to that observed with prospective qMT. Section~\ref{sec:discussion_limitations} discusses limitations related to the report-derived diagnoses, lesion segmentations, and clinical characterization of the data.


For retrospective application to heterogeneous clinical MRI archives, MPF estimates must remain stable across acquisition protocols and scanner systems. Across the multicenter test cohort, mean MPF varied little among the three vendor-by-protocol groups, with between-group CVs of 2.65\% in WM and 5.63\% in GM, despite differences in scanner vendor, sequence implementation, timing parameters, and RF power. The prospective within-subject experiment further supported this protocol robustness: MPF maps estimated from consecutive Spoiled GRE and MPRAGE acquisitions showed good voxel-wise agreement (CCC $=0.888$) and a mean absolute percentage difference of 7.7\% (Figures~\ref{fig:volunteer_maps} and~\ref{fig:volunteer_scatter}).

Acquisition-conditioned AdaIN contributed substantially to this stability. Replacing AdaIN with standard instance normalization substantially increased the between-group CV of MPF, and reduced cross-protocol agreement in the prospective volunteer, with the CCC decreasing from $0.888$ to $0.672$ (Supplementary Materials~\ref{sec:appendix_results_conditioning}). This finding is consistent with other quantitative mapping frameworks showing that conditioning neural networks on sequence parameters improves performance across heterogeneous acquisition settings~\cite{Chen2026,Qiu2024ISMRM}. However, protocol hold-out experiments indicated limited generalization to sequence types absent from the training data (Supplementary Materials~\ref{sec:appendix_domain_shift}). Robust deployment across clinical archives therefore requires representative coverage of the target sequence types during training.

The observed protocol stability is particularly relevant given the ill-posed nature of the parameter-estimation problem. The framework estimates four quantitative maps ($m_0^\mathrm{s}$, $T_1^\mathrm{f}$, $T_2^\mathrm{f}$, and $\mathrm{PD}$) from only three conventional image contrasts ($T_1\mathrm{w}$, $T_2\mathrm{w}$, and FLAIR). Moreover, as discussed in the Methods (Section~\ref{sec:objective}), $T_1^\mathrm{f}$ and $m_0^\mathrm{s}$ exhibit a specific identifiability problem. To constrain the solution space, the model incorporates two regularizers: a WM-specific $T_1^\mathrm{f}$ prior and a soft lower bound on $\mathrm{PD}$. The importance of these regularizers was evaluated by the ablation study (Supplementary Materials~\ref{sec:appendix_ablation_study}). Removing the WM $T_1^\mathrm{f}$ prior caused shifts in $T_1^\mathrm{f}$ and MPF, increased differences between vendor-by-protocol groups, and reduced cross-protocol agreement in the prospective volunteer experiment. Removing the $\mathrm{PD}$ constraint primarily affected the estimated $\mathrm{PD}$ and $T_2^\mathrm{f}$ maps and increased vendor-by-protocol group variation. These findings suggest that the regularization improves the stability and reproducibility of the generated quantitative maps. This is consistent with previous physics-guided self-supervised approaches that used similar regularizers to derive quantitative maps from conventional MRI~\cite{vanLune2026,Qiu2024}. Beyond these regularization constraints, protocol-dependent variation in sequence-averaged RF power ($B_{1,\mathrm{rms}}$) provides additional information that may help distinguish the contributions of $T_1^\mathrm{f}$ and $m_0^\mathrm{s}$ to the observed image intensities. Combining the UMCU and MR-RATE datasets broadened the range of acquisition protocols and corresponding $B_{1,\mathrm{rms}}$ values represented during training (Table~\ref{tab:table_protocol_stacked}).

\subsection{Limitations}
\label{sec:discussion_limitations}
Several limitations should be considered when interpreting our findings. First, the framework relies on simplified descriptions of MT and conventional MRI signal formation. The analytical MT model adopted from Teixeira et al.~\cite{Rui2019} incorporates several assumptions, as described in the Theory section (Section~\ref{sec:theory}). In addition, the Bloch-based signal equations do not account for factors such as $B_1$ inhomogeneity or detailed echo-train dynamics. More comprehensive approaches, including the MT models proposed by~\citet{Graham1997}, and~\citet{Sled2001}, EPG-X~\cite{Malik2018}, and the generalized Bloch model~\cite{Asslnder2024b}, may provide more accurate descriptions of these effects. However, these methods generally require detailed sequence information that is often unavailable in retrospective clinical MRI archives.

The model also relies on fixed tissue parameters and explicit regularization to constrain parameter estimation. We fixed the exchange and semisolid relaxation parameters to literature values reported for healthy WM and applied them to all tissues and subjects. Because these parameters vary across tissue types and may also change with pathology~\cite{Asslnder2024a}, applying fixed healthy-WM values to all tissues and subjects may bias the estimated MPF. Moreover, although the WM-specific $T_1^\mathrm{f}$ prior and soft lower bound on $\mathrm{PD}$ improved model stability, they constrained the range of admissible solutions and made the estimated parameter maps dependent on the selected regularization settings.

In addition to these modeling assumptions, the heterogeneous acquisition geometry of clinical MRI imposed spatial limitations, as previously discussed in~\cite{vanLune2026}. We registered all MRI contrasts to the 2D $T_2\mathrm{w}$ acquisition geometry and resampled them to an in-plane resolution of $1 \times 1$~mm. This procedure reduced the through-plane resolution of the original 3D $T_1\mathrm{w}$ and FLAIR images and removed volumetric detail. Registration errors and partial-volume effects may also propagate into the estimated parameter maps. Moreover, slice-wise processing does not use spatial information from adjacent slices.


The evaluation design also has several limitations. Establishing the absolute accuracy of the MPF maps remains challenging because no universally accepted qMT reference standard exists~\cite{Keenan2019,Kisel2022}. Available qMT methods differ in sequence design, fitting procedures, and the constraints imposed on the two-pool model, which can lead to systematic differences in the estimated MPF~\cite{Kisel2022}. Comparison with a dedicated prospective qMT implementation would therefore establish agreement with that specific methodology, rather than establishing absolute biological accuracy. In the absence of paired qMT data, we evaluated the generated MPF maps through complementary measures of biological plausibility, sensitivity to demyelinating pathology, stability across vendor-by-protocol groups, and prospective cross-protocol reproducibility. Furthermore, these analyses relied on tissue segmentations derived from clinical MRI, which may be inaccurate in the presence of lesions, anatomical abnormalities, or limited image quality. Errors in the WM and GM masks may affect the reported tissue-level MPF estimates, while errors in the automated WM lesion segmentations may affect the lesion analysis. Moreover, we relied primarily on report-derived diagnostic labels to evaluate demyelinating pathology, and FLAIR-hyperintense white-matter lesions are not specific to demyelination. The four clinically confirmed MS UMCU cases supported the findings but did not constitute a systematic validation cohort. These caveats should be considered when interpreting the observed reduction in lesion MPF. Future studies should evaluate the framework in clinically characterized cohorts with expert tissue and lesion annotations.

While the large and heterogeneous clinical dataset enabled evaluation across multiple vendors and protocols, the vendor-by-protocol groups differed in patient demographics. As a result, biological variability across the populations confounded the group comparisons, preventing an independent assessment of the effects of vendor and protocol on MPF estimates. Moreover, all MRIs were acquired at 3~T, and protocol diversity was concentrated mainly in the $T_1\mathrm{w}$ acquisitions; other common acquisitions, including 3D $T_2\mathrm{w}$ and 2D FLAIR, were not represented. Generalizability beyond the scanner systems and acquisition protocols represented in this study remains to be established. In particular, protocol hold-out experiments showed substantial parameter shifts for $T_1\mathrm{w}$ sequence types absent from training (Supplementary Materials~\ref{sec:appendix_domain_shift}); the framework should therefore not be applied to unseen sequence types without retraining or dedicated validation. Finally, the prospective reproducibility experiment included only a single healthy volunteer and should therefore be considered a technical demonstration; larger, dedicated cohorts are required to prospectively assess cross-protocol and cross-vendor reproducibility more comprehensively.

\subsection{Outlook}
\label{sec:discussion_outlook}

By deriving MPF maps directly from routine $T_1\mathrm{w}$, $T_2\mathrm{w}$, and FLAIR scans, the framework enables quantitative MT analysis without dedicated qMT acquisitions. The framework can generate MPF maps from existing multicenter conventional clinical imaging archives, supporting large-scale retrospective biomarker research. Applied prospectively, it can provide MPF as a quantitative biomarker without increasing scan time and may support longitudinal analysis of lesions in patients undergoing repeated clinical MRI. Although the present study focuses on MS, the proposed MPF mapping approach may also enable investigation of macromolecular tissue changes in other neurological conditions. Potential applications include Alzheimer's disease~\cite{Kiefer2009,Mole2020}, Parkinson's disease~\cite{Trujillo2017}, Huntington's disease~\cite{Bourbon-Teles2019}, and myelin damage and repair after stroke~\cite{Khodanovich2021}.

\section{Conclusion}
Physics-guided self-supervised learning enables retrospective estimation of MPF maps from conventional clinical brain MRI without requiring paired qMT reference data. Across heterogeneous protocols, the resulting maps showed good consistency and marked MPF reductions in lesions of patients with demyelinating disease. This highlights the framework's potential for large-scale biomarker research using existing clinical imaging archives. It may also enable prospective MPF mapping from conventional MRI without requiring a dedicated qMT acquisition or additional scan time.

\section*{Acknowledgments}
The Hanarth Fonds for AI in Oncology supported this research. 

\section*{Data availability} 
The UMCU data cannot be shared because of privacy and institutional restrictions. The MR-RATE dataset is available through Hugging Face at \url{https://huggingface.co/datasets/Forithmus/MR-RATE} (accessed September 8, 2026), subject to its access conditions and terms of use. The sessions included in this study were selected according to the eligibility criteria described in the Methods.

Code and model weights are available on GitHub: \url{https://github.com/JelmervanL/MPF-mapping-from-conventional-MRI}.

\clearpage
\appendix
\counterwithin{figure}{section}
\counterwithin{table}{section}
\counterwithin{equation}{section}

\section{MT signal model assumptions}
\label{sec:appendix_mt_approximations}

The following sections provide further justification for the sequence-averaged RF saturation and near-resonance approximations introduced in the Theory section (Section~\ref{sec:theory}), and explain the omission of the sequence-dependent apparent equilibrium magnetization from the forward signal model.

\subsection{Sequence-averaged RF saturation} 
\label{sec:appendix_pulse_rate}

Teixeira et al.~\cite{Rui2019} applied the continuous-wave-equivalent formulation to short-$\mathrm{TR}$ steady-state acquisitions and defined $B_{1,\mathrm{rms}}$ over a single repetition period. We extend this formulation to conventional clinical acquisitions by averaging the cumulative RF power of all excitation, inversion, and refocusing pulses over the entire acquisition. This extension assumes that the combined RF pulse rate is high relative to the free- and semisolid-pool longitudinal relaxation rates ($R_1^\mathrm{f}$ and $R_1^\mathrm{s}$), the exchange rate ($R_x$), and the sequence-averaged semisolid-pool saturation rate ($\langle\overline{W}\rangle$).

The combined RF pulse rate includes pulses applied to the imaged volume and slice-selective pulses applied to other slices, whether on- or off-resonance. Consequently, even acquisitions with a long $\mathrm{TR}$ can contain RF pulses at intervals on the order of milliseconds, primarily because of frequent TSE refocusing pulses. These intervals correspond to combined RF pulse rates on the order of $10^2~\mathrm{s}^{-1}$, which exceed the longitudinal relaxation and exchange rates used in the model ($R_1^\mathrm{f}$, $R_1^\mathrm{s}$, and $R_x < 15~\mathrm{s}^{-1}$). The modeled semisolid-pool saturation rate is also substantially lower than the combined RF pulse rate. For example, Eq.~\eqref{eq:w_simplified} gives $\langle\overline{W}\rangle \approx 6~\mathrm{s}^{-1}$ for $B_{1,\mathrm{rms}}=2.3~\mu\mathrm{T}$ and $T_2^\mathrm{s}=12.5~\mu\mathrm{s}$. The semisolid longitudinal magnetization is therefore expected to change little between successive pulses, allowing its cumulative saturation to be approximated using the sequence-averaged RF power.

\subsection{Near-resonance assumption and apparent equilibrium magnetization}
\label{sec:appendix_lineshape_approx} 

In the continuous-wave-equivalent formulation of Teixeira et al.~\cite{Rui2019}, the semisolid-pool saturation rate depends on the absorption lineshape $G(\Delta)$ evaluated at the RF frequency offset $\Delta$. In this work, we assume a Gaussian lineshape:

\begin{equation}
G(\Delta) = (T_2^\mathrm{s} / \sqrt{2\pi}) \exp[ -(2\pi\Delta T_2^\mathrm{s})^2 / 2 ],
\label{eq:appendix_gaussian_lineshape}
\end{equation}

where $\Delta$ denotes the frequency offset from the water resonance. We assume near-resonance irradiation and therefore approximate the lineshape by its peak value, $G(\Delta)\approx G(0)=T_2^\mathrm{s}/\sqrt{2\pi}$, as used in Eq.~\eqref{eq:w_simplified}. This approximation requires $2\pi|\Delta|T_2^\mathrm{s}\ll1$, or equivalently,

\begin{equation}
|\Delta| \ll 1 / (2\pi T_2^\mathrm{s}) \approx 12.7~\mathrm{kHz} \qquad (\text{for } T_2^\mathrm{s} = 12.5~\mu\mathrm{s}).
\label{eq:delta_limit}
\end{equation}

This near-resonance assumption is generally satisfied for the 3D acquisitions considered in this study, as all RF pulses are applied at or near the water resonance. The assumption is less valid for the multislice 2D $T_2\mathrm{w}$ acquisitions in our dataset. An RF pulse applied to one slice is off-resonant for spins in the other slices. The corresponding frequency offsets increase with interslice distance and may no longer be small relative to $1/(2\pi T_2^\mathrm{s})$, as required by Eq.~\eqref{eq:delta_limit}. Explicitly modeling these contributions would require information about pulse timing, waveforms, and slice selection that is typically unavailable in retrospective clinical MRI archives. We therefore applied the assumption $G(\Delta)\approx G(0)$ to all acquisitions, yielding an analytical saturation model that can be used efficiently during self-supervised network training.


Under the continuous-wave-equivalent approximation, Teixeira et al.~\cite{Rui2019} showed that the longitudinal dynamics of the two-pool MT system can be represented by an effective single-pool model. This model is characterized by an apparent longitudinal relaxation rate, $R_1^{\mathrm{app}}$ (Eq.~\eqref{eq:r1_apparent}), and a sequence-dependent apparent equilibrium magnetization: 

\begin{equation} 
M_0^{\mathrm{app}} = M_0^\mathrm{f} \frac{ R_1^\mathrm{f} + R_x m_0^\mathrm{s} \left( 1- \frac{ R_x(1-m_0^\mathrm{s})+\langle\overline{W}\rangle }{ R_1^\mathrm{s} + R_x(1-m_0^\mathrm{s}) + \langle\overline{W}\rangle } \right) }{ R_1^{\mathrm{app}} }. 
\label{eq:appendix_m0_apparent} \end{equation}

In our forward signal models (Eqs.~\eqref{eq:siggre}--\eqref{eq:sigflair}), we represented $M_0^{\mathrm{app}}$ by a single sequence-independent relative $\mathrm{PD}$ map. This simplification neglects the sequence-dependent modulation of the equilibrium-magnetization contribution by the semisolid-pool saturation rate $\langle\overline{W}\rangle$, as described by $M_0^{\mathrm{app}}$.

This simplification was motivated primarily by the multislice 2D $T_2\mathrm{w}$ acquisitions. These acquisitions have the highest $B_{1,\mathrm{rms}}$ values in the dataset (Table~\ref{tab:table_protocol_stacked}) and would therefore show the largest changes in signal amplitude through $M_0^{\mathrm{app}}$. However, their modeled semisolid-pool saturation rates are also the most uncertain because the near-resonance assumption has limited validity for multislice 2D imaging (as discussed above). Explicitly modeling $M_0^{\mathrm{app}}$ would therefore propagate uncertainty in $\langle\overline{W}\rangle$ (Eq.~\eqref{eq:w_simplified}) directly into the modeled $T_2\mathrm{w}$ signal amplitude and potentially affect the estimated MPF maps. 

By instead approximating the equilibrium signal amplitude using a sequence-independent $\mathrm{PD}$, the influence of $\langle\overline{W}\rangle$ on the modeled $T_2\mathrm{w}$ signal is restricted to longitudinal recovery through $R_1^{\mathrm{app}}$. Because the 2D $T_2\mathrm{w}$ acquisitions have long repetition times ($\mathrm{TR}\geq2779~\mathrm{ms}$; Table~\ref{tab:table_protocol_stacked}), longitudinal recovery is nearly complete over the relevant range of $R_1^{\mathrm{app}}$. Uncertainty in $\langle\overline{W}\rangle$ therefore has only a limited effect on the modeled $T_2\mathrm{w}$ signal through this pathway. The estimated $\mathrm{PD}$ should therefore be interpreted as an effective, sequence-independent proton-density parameter rather than as the free-pool equilibrium magnetization $M_0^\mathrm{f}$ or the sequence-dependent apparent equilibrium magnetization $M_0^{\mathrm{app}}$.

\bibliographystyle{unsrtnat}
\bibliography{references}  

\clearpage

\title{Supplementary Materials of: "Retrospective macromolecular pool fraction mapping from conventional clinical MRI: a physics-guided self-supervised deep learning framework"}
\maketitle

\clearpage
\pagenumbering{arabic}
\setcounter{page}{1}
\renewcommand{\thepage}{S\arabic{page}} 

\setcounter{section}{0}
\renewcommand{\thesection}{S\arabic{section}}



\setcounter{section}{0}
\renewcommand{\thesection}{S\arabic{section}}

\counterwithin{figure}{section}
\counterwithin{table}{section}
\counterwithin{equation}{section}

\renewcommand{\thefigure}{\thesection.\arabic{figure}}
\renewcommand{\thetable}{\thesection.\arabic{table}}
\renewcommand{\theequation}{\thesection.\arabic{equation}}








\section{Characteristics of the UMCU multiple sclerosis subset}
\label{sec:appendix_umcu_ms}

\begin{table}[h]
\centering
\small
\setlength{\tabcolsep}{4pt}
\renewcommand{\arraystretch}{1.15}
\caption{Demographic and acquisition parameters for the four subjects with clinically confirmed MS in the UMCU test set. Subjects A--D correspond to Rows A--D in Figure~\ref{fig:montage_UMCU_ms} in the main manuscript. All examinations were performed on Philips 3~T systems. GRE: Gradient-Echo; TSE: Turbo Spin Echo; IR: Inversion Recovery; MPRAGE: Magnetization-Prepared Rapid Gradient-Echo; NA: Not Applicable; BW: Pixel Bandwidth; FA: Flip Angle; ETL: Echo Train Length.}
\label{tab:umcu_ms_acquisitions}
\begin{tabular*}{\linewidth}{@{\extracolsep{\fill}} l cccc @{}}
\toprule
 & \multicolumn{2}{c}{\textbf{Spoiled GRE Protocol}} & \multicolumn{2}{c}{\textbf{MPRAGE Protocol}} \\
\cmidrule(lr){2-3} \cmidrule(lr){4-5}
\textbf{Parameter} & \textbf{Subject A} & \textbf{Subject B} & \textbf{Subject C} & \textbf{Subject D} \\
\midrule
\multicolumn{5}{l}{\textbf{Demographics \& Hardware}} \\
\midrule
\textbf{Age [years]} & 35 & 51 & 13 & 49 \\
\textbf{Sex} & Male & Male & Male & Female \\
\textbf{Scanner model} & Ingenia & Ingenia Elition X & Ingenia & Ingenia CX \\
\midrule
\multicolumn{5}{l}{$\boldsymbol{T_1\mathrm{w}}$ \textbf{Sequence}} \\
\midrule
\textbf{Sequence} & 3D Spoiled GRE & 3D Spoiled GRE & 3D MPRAGE & 3D MPRAGE \\
\textbf{TR [ms]} & 5.30 & 5.30 & 2500 & 2500 \\
\textbf{TE [ms]} & 2.40 & 2.41 & 3.86 & 4.02 \\
\textbf{TI [ms]} & NA & NA & 1004 & 1004 \\
\textbf{FA [$^\circ$]} & 10 & 10 & 8 & 8 \\
\textbf{\boldmath $B_{1,\mathrm{rms}}$ [$\mu$T]} & 0.93 & 0.93 & 0.73 & 0.73 \\
\textbf{ETL} & NA & NA & NA & NA \\
\textbf{BW [Hz/px]} & 539 & 539 & 190 & 190 \\
\textbf{Rec.\ matrix (in-plane)} & $480 \times 480$ & $480 \times 480$ & $240 \times 240$ & $240 \times 240$ \\
\textbf{Rec.\ matrix (slices)} & 320 & 320 & 190 & 160 \\
\midrule
\multicolumn{5}{l}{$\boldsymbol{T_2\mathrm{w}}$ \textbf{Sequence}} \\
\midrule
\textbf{Sequence} & 2D TSE & 2D TSE & 2D TSE & 2D TSE \\
\textbf{TR [ms]} & 3906 & 4123 & 4340 & 3849 \\
\textbf{TE [ms]} & 80 & 80 & 80 & 80 \\
\textbf{TI [ms]} & NA & NA & NA & NA \\
\textbf{FA [$^\circ$]} & 90 & 90 & 90 & 90 \\
\textbf{\boldmath $B_{1,\mathrm{rms}}$ [$\mu$T]} & 2.34 & 2.34 & 2.34 & 2.34 \\
\textbf{ETL} & 15 & 15 & 15 & 15 \\
\textbf{BW [Hz/px]} & 195 & 195 & 195 & 194 \\
\textbf{Rec.\ matrix (in-plane)} & $512 \times 512$ & $512 \times 512$ & $512 \times 512$ & $512 \times 512$ \\
\textbf{Rec.\ matrix (slices)} & 35 & 37 & 39 & 35 \\
\midrule
\multicolumn{5}{l}{\textbf{FLAIR Sequence}} \\
\midrule
\textbf{Sequence} & 3D TSE (IR) & 3D TSE (IR) & 3D TSE (IR) & 3D TSE (IR) \\
\textbf{TR [ms]} & 4800 & 4800 & 4800 & 4800 \\
\textbf{TE [ms]} & 355.5 & 355.5 & 338.3 & 309.0 \\
\textbf{TI [ms]} & 1650 & 1650 & 1650 & 1650 \\
\textbf{FA [$^\circ$]} & 90 & 90 & 90 & 90 \\
\textbf{\boldmath $B_{1,\mathrm{rms}}$ [$\mu$T]} & 1.00 & 1.00 & 1.00 & 1.00 \\
\textbf{ETL} & 199 & 199 & 199 & 199 \\
\textbf{BW [Hz/px]} & 1075 & 1075 & 1075 & 1250 \\
\textbf{Rec.\ matrix (in-plane)} & $240 \times 240$ & $240 \times 240$ & $240 \times 240$ & $240 \times 240$ \\
\textbf{Rec.\ matrix (slices)} & 267 & 267 & 317 & 267 \\
\bottomrule
\end{tabular*}
\end{table}

\clearpage

\section{TSE echo time correction} 
\label{sec:appendix_tse_echo_time}

Turbo spin-echo (TSE) acquisitions use echo trains that can produce stimulated-echo contributions, causing the effective $T_2$ weighting to differ from that predicted by the nominal echo time. In the forward signal models for the $T_2\mathrm{w}$ and FLAIR acquisitions, we accounted for this difference by replacing the nominal echo time with a contrast-equivalent effective echo time, $\mathrm{TE}_{\mathrm{eff}} = c_{\mathrm{TE}}\,\mathrm{TE}_{\mathrm{nom}}$, where $c_{\mathrm{TE}}$ denotes the echo-time correction factor~\cite{Busse2006}.

For the Philips acquisitions, console-reported contrast-equivalent echo times yielded $c_{\mathrm{TE}}=0.90$ for 2D $T_2\mathrm{w}$ and $c_{\mathrm{TE}}=0.42$ for 3D FLAIR. These factors corresponded to an effective echo time of 72~ms for the nominal $T_2\mathrm{w}$ echo time of 80~ms, and a median effective echo time of approximately 132~ms for the median nominal FLAIR echo time of 314.5~ms.

Equivalent console values were unavailable for the Siemens acquisitions. Correction factors were therefore estimated using extended phase graph (EPG) simulations with representative tissue relaxation times of $T_1=1000$~ms and $T_2=100$~ms. Because the echo spacing and center-echo index were not provided in the metadata, the center of $k$-space was assumed to occur halfway through the echo train. For Siemens 2D $T_2\mathrm{w}$, the constant refocusing flip angles ($119^\circ$--$150^\circ$), echo-train lengths (13--20), and nominal echo times reported in the metadata were simulated directly, yielding correction factors between 0.860 and 0.958 ($\mathrm{TE}_{\mathrm{eff}} = 76.5$--$100.1$~ms).

The Siemens 3D FLAIR acquisitions used a SPACE readout with a proprietary variable refocusing flip-angle schedule that was not exported in the metadata. We approximated this schedule using the pseudo-steady-state target-signal formulation of Busse et al.~\cite{Busse2006}, constrained to a maximum refocusing angle of $120^\circ$. Calibrating the normalized target plateau amplitude to $S_{\mathrm{target}} = 0.292$ reproduced the Philips 3D FLAIR correction factor ($c_{\mathrm{TE}}=0.42$) under representative sequence parameters ($\mathrm{TE}=314.5$~ms, $\mathrm{ETL}=199$). Applying this calibrated schedule to the Siemens acquisitions yielded correction factors of 0.502--0.509, corresponding to a median effective echo time of approximately 196~ms.

\clearpage

\section{Hyperparameter tuning}
\label{sec:hyperparameter_tuning}

Table \ref{tab:hyperparams} outlines the hyperparameter grid search space and the specific values explored during the optimization process to identify the optimal model configuration.

\begin{table}[h]
\centering
\caption{Hyperparameter search space and range of values evaluated during the optimization process.}
\label{tab:hyperparams}
\renewcommand{\arraystretch}{1.15} 
\begin{tabular}{ll} 
\toprule
\textbf{Hyperparameter} & \textbf{Explored values} \\
\midrule
Learning rate & [$10^{-6}$, $10^{-5}$, $10^{-4}$, $10^{-3}$, $10^{-2}$] \\
$\lambda_{\mathrm{PD\text{-}constraint}}$ & [0.1, 0.2, 0.25, 0.50, 0.75, 1.0, 2.0] \\
$\lambda_{T_1\text{-}\mathrm{prior}}$ & [0.0001, 0.001, 0.002, 0.0025, 0.005, 0.01] \\
\bottomrule
\end{tabular}
\end{table}

\clearpage

\section{Prospective volunteer acquisition parameters}
\label{sec:appendix_prospective_acq}

\begin{table}[h]
\centering
\small
\setlength{\tabcolsep}{3.5pt}
\renewcommand{\arraystretch}{1.15}
\caption{Acquisition parameters for the two clinical protocols evaluated in the prospective healthy volunteer on a 3~T Philips Ingenia system. Protocols are designated by their respective $T_1\mathrm{w}$ sequence (Spoiled GRE protocol vs.\ MPRAGE protocol). GRE: Spoiled Gradient-Echo; TSE: Turbo Spin Echo; IR: Inversion Recovery; NA: Not Applicable; BW: Pixel Bandwidth; FA: Flip Angle; ETL: Echo Train Length.}
\label{tab:prospective_acquisitions_info}
\begin{tabular*}{\linewidth}{@{\extracolsep{\fill}} l ccc ccc @{}}
\toprule
 & \multicolumn{3}{c}{\textbf{Spoiled GRE Protocol}} & \multicolumn{3}{c}{\textbf{MPRAGE Protocol}} \\
\cmidrule(lr){2-4} \cmidrule(lr){5-7}
\textbf{Parameter} & $\boldsymbol{T_1\mathrm{w}}$ & $\boldsymbol{T_2\mathrm{w}}$ & \textbf{FLAIR} & $\boldsymbol{T_1\mathrm{w}}$ & $\boldsymbol{T_2\mathrm{w}}$ & \textbf{FLAIR} \\
\midrule
\textbf{Sequence} & 3D Spoiled GRE & 2D TSE & 3D TSE (IR) & 3D MPRAGE & 2D TSE & 3D TSE (IR) \\
\textbf{TR [ms]} & 5.07 & 4340 & 4800 & 2500 & 4990 & 4800 \\
\textbf{TE [ms]} & 2.27 & 80 & 336 & 3.79 & 80 & 337 \\
\textbf{TI [ms]} & NA & NA & 1650 & 1004 & NA & 1650 \\
\textbf{FA [$^\circ$]} & 10 & 90 & 90 & 8 & 90 & 90 \\
\textbf{\boldmath $B_{1,\mathrm{rms}}$ [$\mu$T]} & 0.95 & 2.34 & 1.00 & 0.73 & 2.34 & 1.00 \\
\textbf{ETL} & NA & 15 & 199 & NA & 15 & 199 \\
\textbf{BW [Hz/px]} & 542 & 195 & 1047 & 191 & 198 & 1033 \\
\textbf{Rec.\ matrix (in-plane)} & $560 \times 560$ & $512 \times 512$ & $240 \times 240$ & $256 \times 256$ & $560 \times 560$ & $256 \times 256$ \\
\textbf{Rec.\ matrix (slices)} & 380 & 39 & 317 & 180 & 45 & 300 \\
\bottomrule
\end{tabular*}
\end{table}

\clearpage

\section{Quantitative results for all estimated maps}
\label{sec:appendix_results_qmaps}

\begin{table}[h]
\centering
\small
\setlength{\tabcolsep}{4pt}
\renewcommand{\arraystretch}{1.15}
\caption{Quantitative parameter estimates in white matter (WM) and gray matter (GM) across the clinical test set ($n = 1{,}237$ MRI sessions), stratified by the three vendor-by-protocol groups. Values represent mean $\pm$ standard deviation across MRI sessions. The between-group coefficient of variation (CV) was calculated from the three vendor-by-protocol group means (Philips Spoiled GRE, Philips MPRAGE, and Siemens MPRAGE).}
\label{tab:appendix_reproducibility_scanner_sequence_all}
\begin{tabular*}{\linewidth}{@{\extracolsep{\fill}} l c ccc c @{}}
\toprule
& & \multicolumn{2}{c}{\textbf{Philips (UMCU)}} & \textbf{Siemens (MR-RATE)} & \\
\cmidrule(lr){3-4} \cmidrule(lr){5-5}
\textbf{Metric} & \textbf{Total} & \textbf{Spoiled GRE} & \textbf{MPRAGE} & \textbf{MPRAGE} & \textbf{Between-group} \\
& ($n = 1{,}237$) & ($n = 653$) & ($n = 221$) & ($n = 363$) & \textbf{CV (\%)} \\
\midrule
\multicolumn{6}{l}{\textbf{White Matter (WM)}} \\
\addlinespace[0.2em]
$m_0^\mathrm{s}$      & $0.182 \pm 0.007$ & $0.186 \pm 0.006$ & $0.179 \pm 0.004$ & $0.177 \pm 0.006$ & $2.65$ \\
$T_1^\mathrm{f}$ [ms] & $1982 \pm 58$     & $1967 \pm 60$     & $2017 \pm 64$     & $1987 \pm 40$     & $1.26$ \\
$T_2^\mathrm{f}$  [ms]            & $76.6 \pm 13.1$   & $73.8 \pm 13.7$   & $80.9 \pm 18.9$   & $78.8 \pm 1.2$    & $4.68$ \\
$\mathrm{PD}$ [a.u.]   & $0.671 \pm 0.014$ & $0.673 \pm 0.010$ & $0.664 \pm 0.012$ & $0.672 \pm 0.019$ & $0.74$ \\

\midrule
\multicolumn{6}{l}{\textbf{Gray Matter (GM)}} \\
\addlinespace[0.2em]
$m_0^\mathrm{s}$      & $0.144 \pm 0.009$ & $0.149 \pm 0.005$ & $0.147 \pm 0.005$ & $0.134 \pm 0.006$ & $5.63$ \\
$T_1^\mathrm{f}$ [ms] & $2240 \pm 74$     & $2277 \pm 60$     & $2226 \pm 61$     & $2183 \pm 66$     & $2.12$ \\
$T_2^\mathrm{f}$ [ms]            & $87.4 \pm 13.6$   & $86.2 \pm 14.1$   & $91.9 \pm 19.8$   & $87.0 \pm 3.8$    & $3.52$ \\
$\mathrm{PD}$ [a.u.]   & $0.716 \pm 0.020$ & $0.711 \pm 0.012$ & $0.703 \pm 0.017$ & $0.733 \pm 0.021$ & $2.21$ \\

\bottomrule
\end{tabular*}
\end{table}

\begin{table}[h]
\centering
\small
\setlength{\tabcolsep}{4pt}
\renewcommand{\arraystretch}{1.15}
\caption{Quantitative parameter estimates in demyelinating WM lesions, normal-appearing white matter (NAWM), and normal-appearing gray matter (NAGM) for the MR-RATE demyelinating-disease test subset ($n = 70$ MRI sessions; Siemens Magnetom Vida 3~T). Values represent mean $\pm$ standard deviation across MRI sessions. Relative percentage differences were calculated as $\Delta = (\mu_{\mathrm{lesion}} - \mu_{\mathrm{tissue}})/\mu_{\mathrm{tissue}} \times 100\%$, where the reference tissue was either NAWM or NAGM.}
\label{tab:appendix_mrrate_lesion_quantification}
\begin{tabular*}{\linewidth}{@{\extracolsep{\fill}} l ccccc @{}}
\toprule
\textbf{Metric} & \textbf{NAWM} & \textbf{NAGM} & \textbf{Lesion} & \textbf{$\Delta$ vs.\ NAWM (\%)} & \textbf{$\Delta$ vs.\ NAGM (\%)} \\
\midrule
$m_0^\mathrm{s}$      & $0.176 \pm 0.005$ & $0.133 \pm 0.005$ & $0.113 \pm 0.015$  & $-35.6\%$ & $-15.2\%$ \\
$T_1^\mathrm{f}$ [ms] & $1981 \pm 38$     & $2182 \pm 59$     & $2369 \pm 209$     & $+19.6\%$ & $+8.6\%$  \\
$T_2^\mathrm{f}$ [ms]            & $78.5 \pm 0.8$    & $86.5 \pm 2.1$    & $105.8 \pm 16.2$   & $+34.8\%$ & $+22.3\%$ \\
$\mathrm{PD}$ [a.u.]   & $0.678 \pm 0.015$ & $0.736 \pm 0.018$ & $0.842 \pm 0.045$  & $+24.2\%$ & $+14.4\%$ \\
\bottomrule
\end{tabular*}
\end{table}

\begin{table}[h]
\centering
\small
\setlength{\tabcolsep}{4pt}
\renewcommand{\arraystretch}{1.15}
\caption{Subject-level quantitative parameter estimates in normal-appearing white matter (NAWM), normal-appearing gray matter (NAGM), and demyelinating WM lesions for the four clinically confirmed multiple sclerosis subjects in the UMCU test set acquired on Philips 3~T systems. Values represent voxel-wise mean $\pm$ standard deviation within each tissue compartment for each subject. Subjects A--D correspond to Rows A--D in Figure~\ref{fig:montage_UMCU_ms} and Supplementary Table~\ref{tab:umcu_ms_acquisitions}. Relative differences were calculated separately for each subject as $\Delta = (\mu_{\mathrm{lesion}} - \mu_{\mathrm{tissue}})/\mu_{\mathrm{tissue}} \times 100\%$, where the reference tissue was either NAWM or NAGM.}
\label{tab:appendix_lesion_quantification}
\begin{tabular*}{\linewidth}{@{\extracolsep{\fill}} lll cccc @{}}
\toprule
\textbf{Subject} & \textbf{Protocol} & \textbf{Compartment} & \textbf{\boldmath $m_0^\mathrm{s}$} & \textbf{\boldmath $T_1^\mathrm{f}$ [ms]} & \textbf{\boldmath $T_2^\mathrm{f}$ [ms]} & \textbf{\boldmath $\mathrm{PD}$ [a.u.]} \\
\midrule
\multirow{5}{*}{Subject A} & \multirow{5}{*}{Spoiled GRE} 
  & NAWM                    & $0.193 \pm 0.026$ & $1922 \pm 191$    & $69.0 \pm 6.6$    & $0.656 \pm 0.051$ \\
  & & NAGM                   & $0.150 \pm 0.036$ & $2245 \pm 430$    & $82.8 \pm 21.5$   & $0.703 \pm 0.090$ \\
  & & Lesion          & $0.127 \pm 0.036$ & $2447 \pm 580$    & $114.6 \pm 57.6$  & $0.769 \pm 0.090$ \\
  & & $\Delta$ vs.\ NAWM (\%) & $-34.2\%$         & $+27.3\%$         & $+66.1\%$         & $+17.2\%$ \\
  & & $\Delta$ vs.\ NAGM (\%) & $-15.3\%$         & $+9.0\%$          & $+38.4\%$         & $+9.4\%$ \\
\addlinespace[0.5em]
\multirow{5}{*}{Subject B} & \multirow{5}{*}{Spoiled GRE} 
  & NAWM                    & $0.182 \pm 0.023$ & $1949 \pm 180$    & $68.1 \pm 4.5$    & $0.671 \pm 0.048$ \\
  & & NAGM                   & $0.143 \pm 0.031$ & $2262 \pm 354$    & $78.3 \pm 10.1$   & $0.711 \pm 0.081$ \\
  & & Lesion           & $0.121 \pm 0.038$ & $2520 \pm 599$    & $108.4 \pm 39.1$  & $0.787 \pm 0.100$ \\
  & & $\Delta$ vs.\ NAWM (\%) & $-33.5\%$         & $+29.3\%$         & $+59.2\%$         & $+17.3\%$ \\
  & & $\Delta$ vs.\ NAGM (\%) & $-15.4\%$         & $+11.4\%$         & $+38.4\%$         & $+10.7\%$ \\
\addlinespace[0.5em]
\multirow{5}{*}{Subject C} & \multirow{5}{*}{MPRAGE} 
  & NAWM                    & $0.170 \pm 0.021$ & $1988 \pm 126$    & $71.2 \pm 5.9$    & $0.701 \pm 0.068$ \\
  & & NAGM                   & $0.131 \pm 0.021$ & $2253 \pm 224$    & $83.6 \pm 7.9$    & $0.756 \pm 0.092$ \\
  & & Lesion          & $0.127 \pm 0.029$ & $2332 \pm 473$    & $115.4 \pm 141.5$ & $0.821 \pm 0.102$ \\
  & & $\Delta$ vs.\ NAWM (\%) & $-25.3\%$         & $+17.3\%$         & $+62.1\%$         & $+17.1\%$ \\
  & & $\Delta$ vs.\ NAGM (\%) & $-3.1\%$          & $+3.5\%$          & $+38.0\%$         & $+8.6\%$ \\
\addlinespace[0.5em]
\multirow{5}{*}{Subject D} & \multirow{5}{*}{MPRAGE} 
  & NAWM                    & $0.179 \pm 0.017$ & $1959 \pm 87$     & $70.4 \pm 4.7$    & $0.658 \pm 0.050$ \\
  & & NAGM                   & $0.142 \pm 0.027$ & $2188 \pm 264$    & $81.7 \pm 11.5$   & $0.716 \pm 0.097$ \\
  & & Lesion           & $0.137 \pm 0.034$ & $2299 \pm 520$    & $107.1 \pm 34.1$  & $0.783 \pm 0.123$ \\
  & & $\Delta$ vs.\ NAWM (\%) & $-23.5\%$         & $+17.4\%$         & $+52.1\%$         & $+19.0\%$ \\
  & & $\Delta$ vs.\ NAGM (\%) & $-3.5\%$          & $+5.1\%$          & $+31.1\%$         & $+9.4\%$ \\
\bottomrule
\end{tabular*}
\end{table}

\begin{figure}[h]
    \centering
    \includegraphics[width=\linewidth]{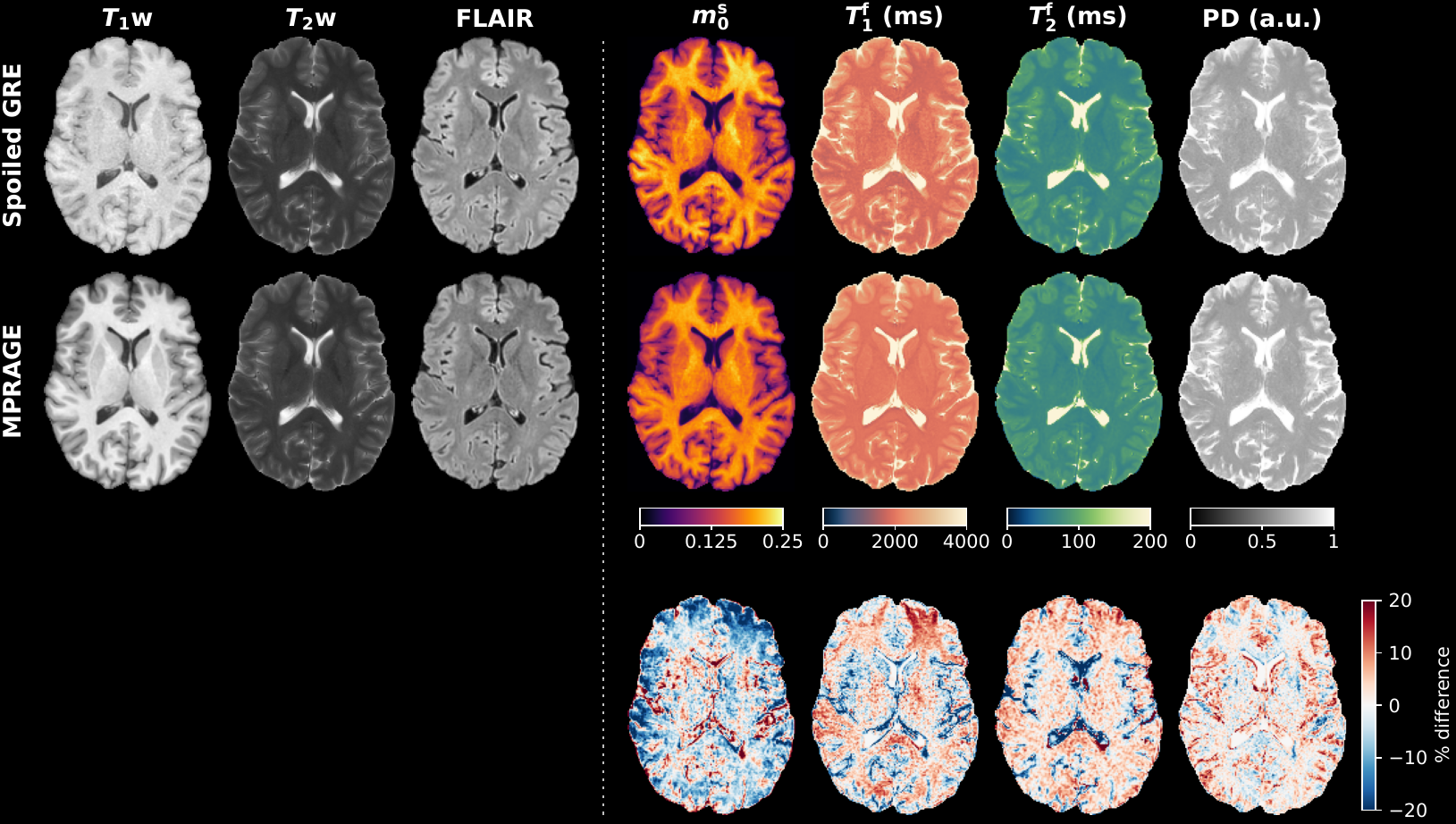}
    \caption{Prospective within-subject reproducibility in a healthy volunteer scanned with two different clinical protocols. Columns show the generated macromolecular pool fraction ($m_0^\mathrm{s}$; MPF), free-pool longitudinal relaxation time ($T_1^\mathrm{f}$), free-pool transverse relaxation time ($T_2^\mathrm{f}$), and relative proton density ($\mathrm{PD}$) maps. The top row corresponds to the Spoiled GRE protocol, the middle row to the MPRAGE protocol, and the bottom row shows voxel-wise percentage difference maps between the two generated quantitative maps. Mean absolute percentage differences across brain parenchyma (WM+GM) were 7.7\% for $m_0^\mathrm{s}$, 7.7\% for $T_1^\mathrm{f}$, 4.8\% for $T_2^\mathrm{f}$, and 4.8\% for $\mathrm{PD}$.}
    \label{fig:appendix_volunteer_slices_full}
\end{figure}

\begin{figure}[h]
    \centering
    \includegraphics[width=\linewidth]{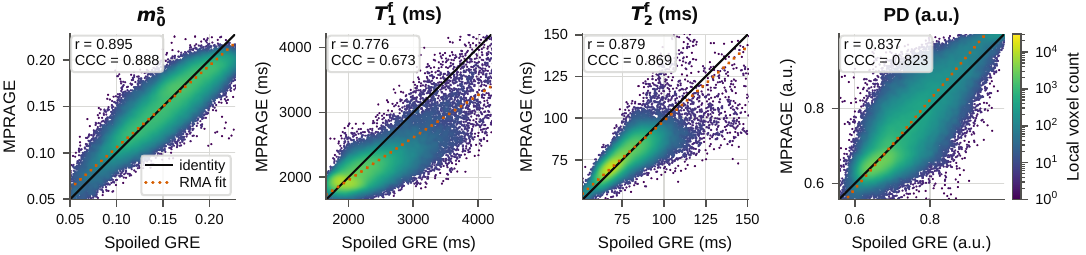}
    \caption{Voxel-wise agreement between quantitative maps estimated from the Spoiled GRE and MPRAGE
    protocols in a healthy volunteer. From left to right, panels show the macromolecular pool fraction
    ($m_0^\mathrm{s}$; MPF), free-pool longitudinal relaxation time ($T_1^\mathrm{f}$), free-pool
    transverse relaxation time ($T_2^\mathrm{f}$), and relative proton density ($\mathrm{PD}$). Each point
    represents a brain parenchymal voxel (WM+GM), colored by local voxel density on a logarithmic scale
    shared by all panels. The solid line denotes identity and the dotted line the reduced major axis (RMA)
    fit. Pearson correlation coefficients ($r$) and concordance correlation coefficients (CCC) are reported
    for each quantitative map to assess cross-protocol reproducibility.}
    \label{fig:appendix_volunteer_scatter_full}
\end{figure}

\clearpage

\section{Impact of acquisition conditioning}
\label{sec:appendix_results_conditioning}


\begin{table*}[h]
\centering
\small
\renewcommand{\arraystretch}{1.15}
\caption{Ablation study evaluating the impact of acquisition-conditioned adaptive instance normalization (AdaIN) on quantitative parameter estimates across the multicenter clinical test set ($n = 1{,}237$ MRI sessions). Values represent mean $\pm$ standard deviation across MRI sessions for white matter (WM) and gray matter (GM). Results are shown for the proposed acquisition-conditioned model (AdaIN) and an otherwise identical unconditioned baseline model (No AdaIN). Group-specific estimates are reported for the three vendor-by-protocol groups (Philips Spoiled GRE, Philips MPRAGE, and Siemens MPRAGE). The between-group coefficient of variation (CV) was calculated from the corresponding groups.}
\label{tab:appendix_adain_tissue_comparison}
\begin{tabular*}{\linewidth}{@{\extracolsep{\fill}} ll c ccc c @{}}
\toprule
& & & \multicolumn{2}{c}{\textbf{Philips (UMCU)}} & \textbf{Siemens (MR-RATE)} & \\
\cmidrule(lr){4-5} \cmidrule(lr){6-6}
\textbf{Parameter} & \textbf{Conditioning} & \textbf{Total} & \textbf{Spoiled GRE} & \textbf{MPRAGE} & \textbf{MPRAGE} & \textbf{Between-group} \\
& & ($n = 1{,}237$) & ($n = 653$) & ($n = 221$) & ($n = 363$) & \textbf{CV (\%)} \\
\midrule
\multicolumn{7}{l}{\textbf{White Matter (WM)}} \\
\addlinespace[0.2em]
$m_0^\mathrm{s}$ & AdaIN & $0.182 \pm 0.007$ & $0.186 \pm 0.006$ & $0.179 \pm 0.004$ & $0.177 \pm 0.006$ & 2.65 \\
& No AdaIN & $0.189 \pm 0.018$ & $0.205 \pm 0.008$ & $0.174 \pm 0.004$ & $0.171 \pm 0.004$ & 10.27 \\
\addlinespace
$T_1^\mathrm{f}$ [ms] & AdaIN & $1982 \pm 58$ & $1967 \pm 60$ & $2017 \pm 64$ & $1987 \pm 40$ & 1.26 \\
& No AdaIN & $2012 \pm 62$ & $2059 \pm 40$ & $1971 \pm 45$ & $1951 \pm 20$ & 2.87 \\
\addlinespace
$T_2^\mathrm{f}$  [ms] & AdaIN & $76.6 \pm 13.1$ & $73.8 \pm 13.7$ & $80.9 \pm 18.9$ & $78.8 \pm 1.2$ & 4.68 \\
& No AdaIN & $75.0 \pm 10.7$ & $72.0 \pm 12.1$ & $76.2 \pm 12.1$ & $79.6 \pm 1.2$ & 5.03 \\
\addlinespace
$\mathrm{PD}$ [a.u.]  & AdaIN & $0.671 \pm 0.014$ & $0.673 \pm 0.010$ & $0.664 \pm 0.012$ & $0.672 \pm 0.019$ & 0.74 \\
& No AdaIN & $0.679 \pm 0.013$ & $0.685 \pm 0.009$ & $0.670 \pm 0.010$ & $0.673 \pm 0.013$ & 1.19 \\
\midrule
\multicolumn{7}{l}{\textbf{Gray Matter (GM)}} \\
\addlinespace[0.2em]
$m_0^\mathrm{s}$ & AdaIN & $0.144 \pm 0.009$ & $0.149 \pm 0.005$ & $0.147 \pm 0.005$ & $0.134 \pm 0.006$ & 5.63 \\
& No AdaIN & $0.146 \pm 0.014$ & $0.158 \pm 0.006$ & $0.138 \pm 0.005$ & $0.129 \pm 0.006$ & 10.37 \\
\addlinespace
$T_1^\mathrm{f}$ [ms] & AdaIN & $2240 \pm 74$ & $2277 \pm 60$ & $2226 \pm 61$ & $2183 \pm 66$ & 2.12 \\
& No AdaIN & $2176 \pm 68$ & $2215 \pm 55$ & $2135 \pm 47$ & $2129 \pm 53$ & 2.21 \\
\addlinespace
$T_2^\mathrm{f}$ [ms] & AdaIN & $87.4 \pm 13.6$ & $86.2 \pm 14.1$ & $91.9 \pm 19.8$ & $87.0 \pm 3.8$ & 3.52 \\
& No AdaIN & $85.1 \pm 11.3$ & $82.9 \pm 13.1$ & $86.5 \pm 12.3$ & $88.1 \pm 4.4$ & 3.12 \\
\addlinespace
$\mathrm{PD}$ [a.u.] & AdaIN & $0.716 \pm 0.020$ & $0.711 \pm 0.012$ & $0.703 \pm 0.017$ & $0.733 \pm 0.021$ & 2.21 \\
& No AdaIN & $0.725 \pm 0.016$ & $0.728 \pm 0.012$ & $0.706 \pm 0.012$ & $0.731 \pm 0.016$ & 1.86 \\
\bottomrule
\end{tabular*}
\end{table*}

\begin{figure}[h]
    \centering
    \includegraphics[width=0.8\linewidth]{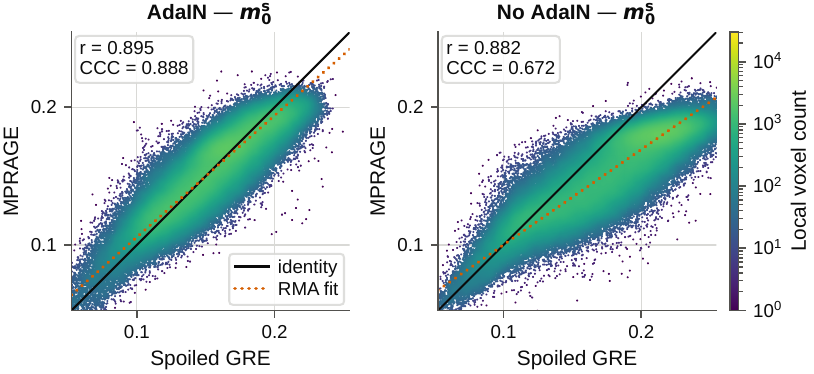}
    \caption{Impact of acquisition conditioning on cross-protocol reproducibility of the generated
    macromolecular pool fraction ($m_0^\mathrm{s}$; MPF) maps in the prospective healthy volunteer.
    Voxel-wise agreement between MPF maps estimated from the Spoiled GRE and MPRAGE protocols is shown for
    the proposed acquisition-conditioned model (AdaIN; left) and an otherwise identical unconditioned
    baseline model (No AdaIN; right). Each point represents a brain parenchymal voxel (WM+GM), colored by
    local voxel density on a logarithmic scale; both panels share the same axis ranges and color scale.
    The solid line denotes identity and the dotted line the reduced major axis (RMA) fit. Pearson
    correlation coefficients ($r$) and concordance correlation coefficients (CCC) quantify cross-protocol
    agreement. Removal of acquisition conditioning reduced reproducibility, resulting in a lower CCC.}
    \label{fig:appendix_volunteer_scatter_adain}
\end{figure}

\clearpage

\section{Ablation of the regularization terms}
\label{sec:appendix_ablation_study}

The training objective in Eq.~\eqref{eq:loss} combines a data-consistency loss with two regularization terms: a soft lower bound on $\mathrm{PD}$ and a prior on white-matter $T_1^\mathrm{f}$. To evaluate their individual contributions, we trained four models differing only in the corresponding loss weights ($\lambda_{\mathrm{PD\text{-}constraint}} \in \{0.5,0\}$ and $\lambda_{T_1\text{-}\mathrm{prior}} \in \{0.0025,0\}$). Each model was evaluated across the clinical test set using the between-group coefficient of variation (CV) and in the prospective healthy volunteer using mean absolute percentage difference (MPD), concordance correlation coefficient (CCC), and Pearson correlation ($r$) (Supplementary Tables~\ref{tab:appendix_reg_ablation_tissue} and~\ref{tab:appendix_reg_ablation_volunteer}).

Removing the white-matter $T_1^\mathrm{f}$ prior primarily affected the estimated $T_1^\mathrm{f}$ and $m_0^\mathrm{s}$ maps. Compared with the full model, these configurations exhibited larger differences between vendor-by-protocol groups and reduced cross-protocol agreement in the prospective volunteer experiment. When both regularization terms were omitted, the estimated $T_1^\mathrm{f}$ values were substantially lower than those obtained with the full model.

Removing the $\mathrm{PD}$ constraint primarily affected $\mathrm{PD}$ and $T_2^\mathrm{f}$, producing larger between-group variation and poorer prospective agreement, as reflected by higher MPD and generally lower CCC. It also shifted $m_0^\mathrm{s}$ to higher values, particularly in WM, although its cross-protocol agreement changed less markedly. Similar trends were observed when both regularization terms were omitted simultaneously.

\begin{table*}[h]
\centering
\small
\renewcommand{\arraystretch}{1.15}
\caption{Regularization ablation: quantitative parameter estimates in white matter (WM) and gray matter (GM) across the clinical test set ($n = 1{,}237$ MRI sessions) for models trained with and without the $\mathrm{PD}$ constraint ($\lambda_{\mathrm{PD\text{-}constraint}} \in \{0.5, 0\}$) and white-matter $T_1^\mathrm{f}$ prior ($\lambda_{T_1\text{-}\mathrm{prior}} \in \{0.0025, 0\}$). Values denote mean $\pm$ standard deviation across sessions. Between-group coefficient of variation (CV) quantifies variation across the three vendor-by-protocol group means (Philips Spoiled GRE, Philips MPRAGE, and Siemens MPRAGE).}
\label{tab:appendix_reg_ablation_tissue}
\begin{tabular*}{\linewidth}{@{\extracolsep{\fill}} ll c ccc c @{}}
\toprule
& & & \multicolumn{2}{c}{\textbf{Philips (UMCU)}} & \textbf{Siemens (MR-RATE)} & \\
\cmidrule(lr){4-5} \cmidrule(lr){6-6}
\textbf{Parameter} & \textbf{Configuration} & \textbf{Total} & \textbf{Spoiled GRE} & \textbf{MPRAGE} & \textbf{MPRAGE} & \textbf{Between-group} \\
& & ($n = 1{,}237$) & ($n = 653$) & ($n = 221$) & ($n = 363$) & \textbf{CV (\%)} \\
\midrule
\multicolumn{7}{l}{\textbf{White Matter (WM)}} \\
\addlinespace[0.2em]
$m_0^\mathrm{s}$ & Full model & $0.182 \pm 0.007$ & $0.186 \pm 0.006$ & $0.179 \pm 0.004$ & $0.177 \pm 0.006$ & 2.65 \\
& No $T_1^\mathrm{f}$ prior & $0.177 \pm 0.013$ & $0.165 \pm 0.005$ & $0.187 \pm 0.003$ & $0.192 \pm 0.004$ & 7.96 \\
& No PD constraint & $0.236 \pm 0.012$ & $0.230 \pm 0.011$ & $0.239 \pm 0.009$ & $0.245 \pm 0.009$ & 3.26 \\
& Neither & $0.169 \pm 0.011$ & $0.159 \pm 0.005$ & $0.179 \pm 0.004$ & $0.180 \pm 0.004$ & 6.89 \\
\addlinespace
$T_1^\mathrm{f}$ [ms] & Full model & $1982 \pm 58$ & $1967 \pm 60$ & $2017 \pm 64$ & $1987 \pm 40$ & 1.26 \\
& No $T_1^\mathrm{f}$ prior & $1866 \pm 284$ & $1611 \pm 94$ & $2110 \pm 92$ & $2177 \pm 65$ & 15.71 \\
& No PD constraint & $1935 \pm 52$ & $1935 \pm 51$ & $1966 \pm 52$ & $1915 \pm 44$ & 1.33 \\
& Neither & $1301 \pm 115$ & $1293 \pm 111$ & $1381 \pm 100$ & $1267 \pm 107$ & 4.54 \\
\addlinespace
$T_2^\mathrm{f}$ [ms] & Full model & $76.6 \pm 13.1$ & $73.8 \pm 13.7$ & $80.9 \pm 18.9$ & $78.8 \pm 1.2$ & 4.68 \\
& No $T_1^\mathrm{f}$ prior & $77.5 \pm 12.4$ & $75.0 \pm 13.1$ & $80.7 \pm 17.6$ & $80.0 \pm 1.2$ & 3.93 \\
& No PD constraint & $86.0 \pm 13.6$ & $82.2 \pm 14.0$ & $96.6 \pm 17.3$ & $86.2 \pm 1.1$ & 8.41 \\
& Neither & $86.2 \pm 13.8$ & $82.3 \pm 14.3$ & $95.9 \pm 17.6$ & $87.5 \pm 1.2$ & 7.78 \\
\addlinespace
$\mathrm{PD}$ [a.u.] & Full model & $0.671 \pm 0.014$ & $0.673 \pm 0.010$ & $0.664 \pm 0.012$ & $0.672 \pm 0.019$ & 0.74 \\
& No $T_1^\mathrm{f}$ prior & $0.664 \pm 0.013$ & $0.667 \pm 0.011$ & $0.660 \pm 0.013$ & $0.661 \pm 0.017$ & 0.52 \\
& No PD constraint & $0.541 \pm 0.045$ & $0.580 \pm 0.016$ & $0.503 \pm 0.018$ & $0.495 \pm 0.019$ & 8.94 \\
& Neither & $0.537 \pm 0.045$ & $0.576 \pm 0.018$ & $0.506 \pm 0.017$ & $0.487 \pm 0.020$ & 8.91 \\
\midrule
\multicolumn{7}{l}{\textbf{Gray Matter (GM)}} \\
\addlinespace[0.2em]
$m_0^\mathrm{s}$ & Full model & $0.144 \pm 0.009$ & $0.149 \pm 0.005$ & $0.147 \pm 0.005$ & $0.134 \pm 0.006$ & 5.63 \\
& No $T_1^\mathrm{f}$ prior & $0.153 \pm 0.009$ & $0.147 \pm 0.005$ & $0.164 \pm 0.004$ & $0.159 \pm 0.005$ & 5.73 \\
& No PD constraint & $0.177 \pm 0.011$ & $0.176 \pm 0.010$ & $0.185 \pm 0.010$ & $0.173 \pm 0.011$ & 3.52 \\
& Neither & $0.151 \pm 0.008$ & $0.145 \pm 0.005$ & $0.163 \pm 0.003$ & $0.156 \pm 0.004$ & 5.64 \\
\addlinespace
$T_1^\mathrm{f}$ [ms] & Full model & $2240 \pm 74$ & $2277 \pm 60$ & $2226 \pm 61$ & $2183 \pm 66$ & 2.12 \\
& No $T_1^\mathrm{f}$ prior & $2414 \pm 233$ & $2214 \pm 97$ & $2577 \pm 86$ & $2674 \pm 90$ & 9.76 \\
& No PD constraint & $2176 \pm 80$ & $2219 \pm 62$ & $2192 \pm 48$ & $2090 \pm 53$ & 3.14 \\
& Neither & $1934 \pm 147$ & $1916 \pm 137$ & $1985 \pm 125$ & $1936 \pm 167$ & 1.83 \\
\addlinespace
$T_2^\mathrm{f}$ [ms] & Full model & $87.4 \pm 13.6$ & $86.2 \pm 14.1$ & $91.9 \pm 19.8$ & $87.0 \pm 3.8$ & 3.52 \\
& No $T_1^\mathrm{f}$ prior & $87.9 \pm 12.6$ & $87.0 \pm 13.3$ & $90.9 \pm 18.3$ & $87.7 \pm 3.4$ & 2.35 \\
& No PD constraint & $96.2 \pm 14.1$ & $94.6 \pm 14.1$ & $108.4 \pm 17.8$ & $91.5 \pm 2.6$ & 9.14 \\
& Neither & $96.0 \pm 13.9$ & $94.3 \pm 14.1$ & $106.8 \pm 18.3$ & $92.5 \pm 2.7$ & 7.99 \\
\addlinespace
$\mathrm{PD}$ [a.u.] & Full model & $0.716 \pm 0.020$ & $0.711 \pm 0.012$ & $0.703 \pm 0.017$ & $0.733 \pm 0.021$ & 2.21 \\
& No $T_1^\mathrm{f}$ prior & $0.710 \pm 0.018$ & $0.708 \pm 0.012$ & $0.699 \pm 0.018$ & $0.722 \pm 0.019$ & 1.65 \\
& No PD constraint & $0.603 \pm 0.035$ & $0.629 \pm 0.018$ & $0.561 \pm 0.022$ & $0.581 \pm 0.025$ & 5.94 \\
& Neither & $0.601 \pm 0.035$ & $0.626 \pm 0.019$ & $0.566 \pm 0.023$ & $0.577 \pm 0.027$ & 5.43 \\
\bottomrule
\end{tabular*}
\end{table*}

\begin{table}[h]
\centering
\small
\renewcommand{\arraystretch}{1.15}
\caption{Regularization ablation: prospective cross-protocol reproducibility of quantitative maps in a healthy volunteer scanned with the Spoiled GRE and MPRAGE protocols. Metrics evaluate voxel-wise agreement across brain parenchyma (WM+GM). MPD: mean absolute percentage difference; CCC: concordance correlation coefficient; $r$: Pearson correlation coefficient.}
\label{tab:appendix_reg_ablation_volunteer}
\begin{tabular*}{\linewidth}{@{\extracolsep{\fill}} ll ccc @{}}
\toprule
\textbf{Parameter} & \textbf{Configuration} & \textbf{MPD (\%)} & \textbf{CCC} & \textbf{$r$} \\
\midrule
$m_0^\mathrm{s}$ & Full model & 7.67 & 0.888 & 0.895 \\
& No $T_1^\mathrm{f}$ prior & 14.60 & 0.473 & 0.793 \\
& No PD constraint & 9.56 & 0.881 & 0.892 \\
& Neither & 12.14 & 0.419 & 0.805 \\
\addlinespace
$T_1^\mathrm{f}$ [ms] & Full model & 7.70 & 0.673 & 0.776 \\
& No $T_1^\mathrm{f}$ prior & 19.57 & 0.722 & 0.898 \\
& No PD constraint & 6.67 & 0.711 & 0.796 \\
& Neither & 12.04 & 0.901 & 0.907 \\
\addlinespace
$T_2^\mathrm{f}$ [ms] & Full model & 4.83 & 0.869 & 0.879 \\
& No $T_1^\mathrm{f}$ prior & 4.77 & 0.864 & 0.881 \\
& No PD constraint & 8.65 & 0.661 & 0.777 \\
& Neither & 10.10 & 0.587 & 0.754 \\
\addlinespace
$\mathrm{PD}$ [a.u.] & Full model & 4.83 & 0.823 & 0.837 \\
& No $T_1^\mathrm{f}$ prior & 4.81 & 0.830 & 0.841 \\
& No PD constraint & 11.31 & 0.697 & 0.842 \\
& Neither & 11.49 & 0.696 & 0.851 \\
\bottomrule
\end{tabular*}
\end{table}

\clearpage

\section{Generalization to unseen \texorpdfstring{$T_1\mathrm{w}$}{T1w} protocols}
\label{sec:appendix_domain_shift}

To evaluate whether the framework generalizes to vendor-by-protocol groups not represented during training, we performed two protocol hold-out experiments based on the $T_1\mathrm{w}$ sequence used to define these groups in the main analysis. We trained one model exclusively on sessions from the Philips and Siemens MPRAGE groups, withholding the Philips Spoiled GRE group. We trained a second model exclusively on sessions from the Philips Spoiled GRE group, withholding both MPRAGE groups. Both protocol-restricted models used the same network architecture, loss weights, and optimization schedule as the pooled model (presented in the main paper) and differed only in the training data provided to them. We evaluated all models on the complete clinical test set ($n = 1{,}237$ sessions). Table~\ref{tab:appendix_domain_shift_summary} summarizes the corresponding training subsets.

\begin{table}[h]
\centering
\small
\setlength{\tabcolsep}{6pt}
\renewcommand{\arraystretch}{1.15}
\caption{Overview of training subsets evaluated in the protocol hold-out experiments. All models were evaluated on the complete clinical test set ($n = 1{,}237$ sessions).}
\label{tab:appendix_domain_shift_summary}
\begin{tabular*}{\linewidth}{@{\extracolsep{\fill}} lll c @{}}
\toprule
\textbf{Configuration} & \textbf{Withheld protocol} & \textbf{Included protocol(s)} & \textbf{Training sessions} \\
\midrule
Pooled (full model) & None             & Spoiled GRE + MPRAGE & 6,084 \\
MPRAGE-only         & Spoiled GRE      & MPRAGE               & 2,773 \\
Spoiled GRE-only    & MPRAGE           & Spoiled GRE          & 3,311\\
\bottomrule
\end{tabular*}
\end{table}

Withholding a protocol during training substantially shifted the parameter estimates for the corresponding out-of-domain vendor-by-protocol groups, indicated by $^{\dagger}$ in Table~\ref{tab:appendix_domain_shift_qvalues}. When applied to the unseen Philips Spoiled GRE group, the MPRAGE-only model produced a mean white-matter MPF of $0.126$, compared with approximately $0.18$ in the two in-domain MPRAGE groups. White-matter $T_1^\mathrm{f}$ and $T_2^\mathrm{f}$ also shifted to markedly higher values. Accordingly, the between-group CV of white-matter MPF increased from $2.65\%$ for the pooled model to $19.71\%$. 

Conversely, when applied to the unseen MPRAGE groups, the Spoiled-GRE-only model produced lower white-matter MPF values and markedly higher white-matter $T_1^\mathrm{f}$ and $T_2^\mathrm{f}$ values than in the in-domain Spoiled GRE group. Mean white-matter MPF decreased from $0.177$ in the Spoiled GRE group to approximately $0.15$ in both MPRAGE groups, increasing the between-group CV to $11.06\%$.

Despite differences in scanner vendor, the MPRAGE-only model produced similar estimates for the Philips and Siemens MPRAGE groups. In contrast, both protocol-restricted models produced substantial parameter shifts when applied to the withheld sequence regime. These findings suggest that stable estimation requires the relevant acquisition protocols to be represented during training.

\begin{table*}[h]
\centering
\footnotesize
\setlength{\tabcolsep}{4pt}
\renewcommand{\arraystretch}{1.15}
\caption{Quantitative parameter estimates in white matter (WM) and gray matter (GM) across the clinical test set ($n = 1{,}237$ MRI sessions) for models trained on protocol-restricted subsets. Values denote mean $\pm$ standard deviation across sessions. $^{\dagger}$ indicates an out-of-domain evaluation where the corresponding protocol was withheld from the training set. Between-group coefficient of variation (CV) was calculated across the three vendor-by-protocol group means.}
\label{tab:appendix_domain_shift_qvalues}
\begin{tabular*}{\linewidth}{@{\extracolsep{\fill}} ll ccc c @{}}
\toprule
& & \multicolumn{2}{c}{\textbf{Philips (UMCU)}} & \textbf{Siemens (MR-RATE)} & \\
\cmidrule(lr){3-4} \cmidrule(lr){5-5}
\textbf{Parameter} & \textbf{Training data} & \textbf{Spoiled GRE} ($n = 653$) & \textbf{MPRAGE} ($n = 221$) & \textbf{MPRAGE} ($n = 363$) & \textbf{Between-group} \\ & & & & & \textbf{CV (\%)} \\
\midrule
\multicolumn{6}{l}{\textbf{White Matter (WM)}} \\
\addlinespace[0.2em]
$m_0^\mathrm{s}$ & Pooled (full model) & $0.186 \pm 0.006$ & $0.179 \pm 0.004$ & $0.177 \pm 0.006$ & 2.65 \\
 & MPRAGE only          & $0.126 \pm 0.004^{\dagger}$ & $0.182 \pm 0.005$ & $0.183 \pm 0.006$ & 19.71 \\
 & Spoiled GRE only     & $0.177 \pm 0.007$ & $0.146 \pm 0.007^{\dagger}$ & $0.148 \pm 0.008^{\dagger}$ & 11.06 \\
\addlinespace
$T_1^\mathrm{f}$ [ms] & Pooled (full model) & $1967 \pm 60$ & $2017 \pm 64$ & $1987 \pm 40$ & 1.26 \\
 & MPRAGE only          & $2240 \pm 48^{\dagger}$ & $2025 \pm 72$ & $2008 \pm 52$ & 6.19 \\
 & Spoiled GRE only     & $1996 \pm 55$ & $3579 \pm 87^{\dagger}$ & $3614 \pm 82^{\dagger}$ & 30.18 \\
\addlinespace
$T_2^\mathrm{f}$ [ms] & Pooled (full model) & $73.8 \pm 13.7$ & $80.9 \pm 18.9$ & $78.8 \pm 1.2$ & 4.68 \\
 & MPRAGE only          & $120.6 \pm 7.1^{\dagger}$ & $78.2 \pm 14.8$ & $79.2 \pm 1.0$ & 26.09 \\
 & Spoiled GRE only     & $77.2 \pm 13.8$ & $118.4 \pm 20.0^{\dagger}$ & $115.1 \pm 11.4^{\dagger}$ & 22.09 \\
\addlinespace
$\mathrm{PD}$ [a.u.] & Pooled (full model) & $0.673 \pm 0.010$ & $0.664 \pm 0.012$ & $0.672 \pm 0.019$ & 0.74 \\
 & MPRAGE only          & $0.680 \pm 0.018^{\dagger}$ & $0.660 \pm 0.013$ & $0.657 \pm 0.015$ & 1.84 \\
 & Spoiled GRE only     & $0.666 \pm 0.010$ & $0.679 \pm 0.033^{\dagger}$ & $0.682 \pm 0.034^{\dagger}$ & 1.29 \\
\midrule
\multicolumn{6}{l}{\textbf{Gray Matter (GM)}} \\
\addlinespace[0.2em]
$m_0^\mathrm{s}$ & Pooled (full model) & $0.149 \pm 0.005$ & $0.147 \pm 0.005$ & $0.134 \pm 0.006$ & 5.63 \\
 & MPRAGE only          & $0.141 \pm 0.004^{\dagger}$ & $0.152 \pm 0.006$ & $0.144 \pm 0.006$ & 3.83 \\
 & Spoiled GRE only     & $0.143 \pm 0.006$ & $0.143 \pm 0.006^{\dagger}$ & $0.147 \pm 0.009^{\dagger}$ & 1.77 \\
\addlinespace
$T_1^\mathrm{f}$ [ms] & Pooled (full model) & $2277 \pm 60$ & $2226 \pm 61$ & $2183 \pm 66$ & 2.12 \\
 & MPRAGE only          & $2254 \pm 57^{\dagger}$ & $2273 \pm 71$ & $2279 \pm 58$ & 0.58 \\
 & Spoiled GRE only     & $2319 \pm 56$ & $3149 \pm 135^{\dagger}$ & $3121 \pm 152^{\dagger}$ & 16.46 \\
\addlinespace
$T_2^\mathrm{f}$ [ms] & Pooled (full model) & $86.2 \pm 14.1$ & $91.9 \pm 19.8$ & $87.0 \pm 3.8$ & 3.52 \\
 & MPRAGE only          & $125.0 \pm 4.7^{\dagger}$ & $88.7 \pm 15.4$ & $86.8 \pm 2.4$ & 21.48 \\
 & Spoiled GRE only     & $91.5 \pm 15.0$ & $130.8 \pm 21.2^{\dagger}$ & $142.1 \pm 23.1^{\dagger}$ & 21.88 \\
\addlinespace
$\mathrm{PD}$ [a.u.] & Pooled (full model) & $0.711 \pm 0.012$ & $0.703 \pm 0.017$ & $0.733 \pm 0.021$ & 2.21 \\
 & MPRAGE only          & $0.652 \pm 0.013^{\dagger}$ & $0.693 \pm 0.020$ & $0.712 \pm 0.019$ & 4.44 \\
 & Spoiled GRE only     & $0.698 \pm 0.012$ & $0.741 \pm 0.025^{\dagger}$ & $0.767 \pm 0.022^{\dagger}$ & 4.75 \\
\bottomrule
\end{tabular*}
\end{table*}

\end{document}